\documentstyle[plenum,epsfig,12pt]{book}
\newcommand{\beq}{\begin{equation}}
\newcommand{\eeq}{\end{equation}}
\newcommand{\bea}{\begin{eqnarray}}
\newcommand{\eea}{\end{eqnarray}}
\newcommand{\half}{{\scriptstyle{{1\over 2}}}}

\newcommand{\site}[1]{\refnote{\cite{#1}}}
\def\pslash{\not{{\hskip-.08cm} p}}
\def\kslash{\not{{\hskip-.08cm} k}}

\def\eg{{\it e.g.}}
\def\ie{{\it i.e.}}
\def\order{{\cal O}}

\def\abinit{{\it ab initio}}
\def\lt{{\mathaccent "7E \lambda}}
\def\lzero{{\Lambda_0}}
\def\lone{{\Lambda_1}}

\def\Lone{{\Lambda_1^2 \over P^+}}
\def\Lam{{\Lambda^2 \over P^+}}
\def\dqt{d\tilde{q}}
\def\dkt{d\tilde{k}}
\def\Dslash{\not{{\hskip-.12cm} D}}
\def\psit{{\tilde{\psi}}}
\def\At{{\tilde{A}}}

\def\h0{{\cal H}^\Lambda_0}
\def\v{{\it v}^\Lambda}
\def\H{{\cal H}^\Lambda}
\def\V{{\cal V}^\Lambda}
\begin{document}

\chapter{LIGHT-FRONT QCD: A CONSTITUENT PICTURE OF HADRONS}

\author{Robert J. Perry}

\affiliation{Department of Physics, The Ohio State University\\
Columbus, Ohio, 43210, USA}

\section{MOTIVATION AND STRATEGY}

We seek to derive the structure of hadrons from the fundamental theory
of the strong interaction, QCD.  Our work is founded on the
hypothesis that a constituent approximation can be {\it derived} from
QCD, so that a relatively small number of quark {\it and gluon}
degrees of freedom need be explicitly included in the state vectors
for low-lying hadrons.\site{nonpert} To obtain a constituent picture,
we use a Hamiltonian approach in light-front coordinates.\site{dirac}

I do not believe that light-front Hamiltonian field theory is
extremely useful for the study of low energy QCD unless a constituent
approximation can be made, and I do not believe such an approximation
is possible unless cutoffs that {it violate} manifest gauge invariance
and covariance are employed.  Such cutoffs {\it inevitably} lead to
relevant and marginal effective interactions ({\it i.e.}, counterterms)
that contain functions of longitudinal momenta.\site{coupcoh,perryrg}
It is {it not} possible to renormalize light-front Hamiltonians in any
useful manner without developing a renormalization procedure that can
produce these  non-canonical counterterms.

The line of investigation I discuss has been developed by a
small group of theorists who are working or have worked at Ohio State
University and Warsaw University.\refnote{1,3-19}
Ken Wilson provided the initial
impetus for this work, and at a very early stage outlined much of the
basic strategy we employ.\site{wilson90}

I make no attempt to provide enough details to allow the reader to
start doing light-front calculations. The introductory article by
Harindranath\site{hari96a} is helpful in this regard. An earlier
version of these lectures\site{brazil} also provides many more
details.

\subsection{A Constituent Approximation Depends on Tailored
Renormalization}

If it is possible to derive a constituent approximation from QCD, we
can formulate the hadronic bound state problem as a set of coupled
few-body problems. We obtain the states and eigenvalues by solving
\begin{equation}
H_\Lambda \mid \Psi_\Lambda\rangle = E \mid \Psi_\Lambda \rangle,
\end{equation}
where,
\begin{equation}
\mid \Psi_\Lambda \rangle = \phi^\Lambda_{q\bar{q}} \mid q\bar{q}
\rangle + \phi^\Lambda_{q\bar{q}g} \mid q\bar{q}g \rangle +
\cdot\cdot\cdot.
\end{equation}
where I use shorthand notation for the Fock space components
of the state.  The full state vector includes an infinite number of
components, and in a constituent approximation we truncate this
series.  We derive the Hamiltonian from QCD, so we must allow for the
possibility of constituent gluons.  I have indicated that the
Hamiltonian and the state both depend on a cutoff, $\Lambda$, which
is critical for the approximation.

This approach has no chance of working without a renormalization scheme
{\it tailored to it.}  Much of our work has focused on the
development of such a renormalization scheme.\refnote{3,4,17-19}
In order to understand
the constraints that have driven this development, seriously consider
under what conditions it might be possible to truncate the above
series without making an arbitrarily large error in the eigenvalue. 
I focus on the eigenvalue, because it will certainly not be possible
to approximate all observable properties of hadrons (\eg, wee parton
structure functions) this way.

For this approximation to be valid, {\it all} many-body states must
approximately decouple from the dominant few-body components.  We know
that even in perturbation theory, high energy many-body states do not
simply decouple from few-body states.  In fact, the errors from simply
discarding high energy states are infinite.  In second-order
perturbation theory, for example, high energy photons contribute an
arbitrarily large shift to the mass of an electron.  This
second-order effect is illustrated in Figure 1, and the precise
interpretation for this light-front time-ordered diagram will be
given below.  The solution to this problem is well-known,
renormalization.  We {\it must} use renormalization to move the
effects of high energy components in the state to effective
interactions\footnote {These include one-body operators that modify
the free dispersion relations.} in the Hamiltonian.

\begin{figure}
\begin{center}
\epsfig{file=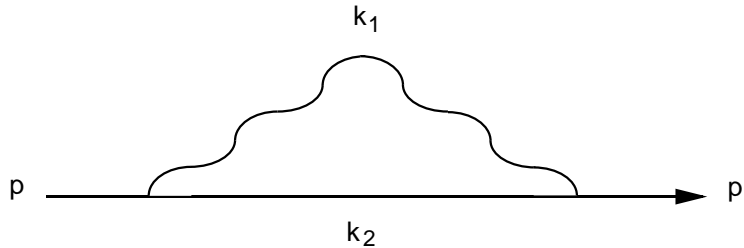,width=12cm}
\end{center} 
\caption{The second-order shift in the self-energy of a bare electron
due to mixing with electron-photon states.}
\label{fig:fig1}
\end{figure}

It is difficult to see how a constituent approximation can emerge
using any regularization scheme that does not employ a cutoff that
either removes degrees of freedom or removes direct couplings between
degrees of freedom.  A Pauli-Villars ``cutoff," for example,
drastically increases the size of Fock space and destroys the
hermiticity of the Hamiltonian.

In the best case scenario we expect the cutoff to act like a
resolution.  If the cutoff is increased to an arbitrarily large value,
the resolution increases and instead of seeing a few constituents we
resolve the substructure of the constituents and the few-body
approximation breaks down. As the cutoff is lowered, this
substructure is removed from the state vectors, and the
renormalization procedure replaces it with effective interactions in
the Hamiltonian.  Any ``cutoff" that does not remove this
substructure from the states is of no use to us.  

This point is well-illustrated by the QED calculations discussed
below\site{brazil,BJ97a,BJ97c}$\!.$ There is a window into which the
cutoff must be lowered for the constituent approximation to work.  If
the cutoff is too large, atomic states must explicitly include
photons.  After the cutoff is lowered to a value that can be
self-consistently determined {\it a-posteriori}, photons are removed
from the states and replaced by the Coulomb interaction and
relativistic corrections.  The cutoff cannot be lowered too far using
a perturbative renormalization group, hence the window.

Thus, if we remove high energy degrees of freedom, or coupling to
high energy degrees of freedom, we should encounter
self-energy shifts leading to effective one-body operators,
vertex corrections leading to effective vertices, and 
exchange effects leading to explicit many-body interactions not found
in the canonical Hamiltonian.  We naively expect these operators to be
local when acting on low energy states,  because simple uncertainty
principle arguments indicate that high energy virtual particles cannot
propagate very far.  Unfortunately this expectation is indeed naive,
and at best we can hope to maintain transverse
locality.\site{perryrg} I will elaborate on this point below. The
study of perturbation theory with the cutoffs we must employ makes it
clear that it is {\it not enough} to adjust the canonical couplings
and masses to renormalize the theory.  It is not possible to make
significant progress towards solving light-front QCD without fully
appreciating this point.

Low energy many-body states do not typically decouple from low energy
few-body states.  The worst of these low energy many-body states is
the vacuum.  This is what drives us to use light-front
coordinates.\site{dirac} Figure 2 shows a pair of particles being
produced out of the vacuum in equal-time coordinates $t$ and $z$.  The
transverse components $x$ and $y$ are not shown, because they are the
same in equal-time and light-front coordinates.  The figure also
shows light-front time,
\begin{equation}
x^+=t+z\;, 
\end{equation}
and the light-front longitudinal spatial coordinate,
\begin{equation}
x^-=t-z\;.
\end{equation}

\begin{figure}
\begin{center}
\epsfig{file=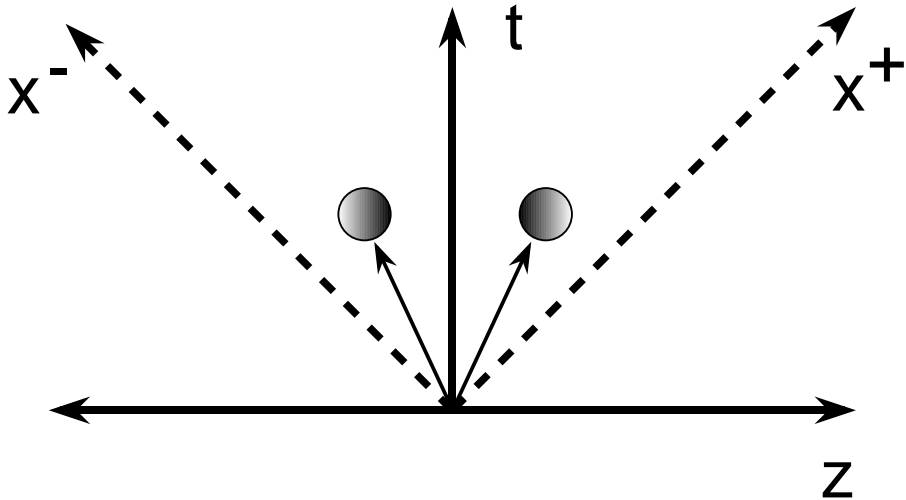,width=12cm}
\end{center} 
\caption{Light-front coordinates. Light-front `time' is $x^+=t+z$, and
the light-front longitudinal spatial coordinate is $x^-=t-z$.}
\label{fig:fig2}
\end{figure}

In equal-time coordinates it is kinematically possible for virtual
pairs to be produced from the vacuum (although relevant interactions
actually produce three or more particles from the vacuum), as long as
their momenta sum to zero so that three-momentum is conserved. 
Because of this, the state vector for a proton includes an arbitrarily
large number of particles that are disconnected from the proton.  The
only constraint imposed by relativity is that particle velocities be
less than or equal to that of light.

In light-front coordinates, however, we see that all allowed
trajectories lie in the first quadrant.  In other words,
light-front longitudinal momentum, $p^+$ (conjugate to $x^-$
since
$a\cdot b=\half (a^+ b^- + a^- b^+) - {\bf a}_\perp \cdot {\bf
b}_\perp$), is always positive,
\begin{equation}
p^+ \ge 0 \;.
\end{equation}
We exclude $p^+=0$, forcing the vacuum to be trivial because it is
the only state with $p^+=0$.  Moreover, the light-front energy
of a free particle of mass $m$ is
\begin{equation}
p^-={{\bf p}_\perp^2+m^2 \over p^+} \;.
\end{equation}
This implies that all free particles with zero longitudinal momentum
have infinite energy, unless their mass and transverse momentum are
identically zero. Replacing such particles with effective
interactions should be reasonable.
\begin{itemize}
\item{Is the vacuum really trivial?}
\item{What about confinement?}
\item{What about chiral symmetry breaking?}
\item{What about instantons?}
\item{What about the job security of theorists who study the vacuum?}
\end{itemize}

The question of how one should treat ``zero modes," degrees of freedom
(which may be constrained) with identically zero longitudinal
momentum, divides the light-front community.  Our attitude is that
explicitly including zero modes defeats the purpose of using
light-front coordinates, and we do not believe that significant
progress will be made in this direction, at least not in $3+1$
dimensions.

The vacuum in our formalism is trivial.  We are forced to work in the
``hidden symmetry phase" of the theory, and to introduce effective
interactions that reproduce all effects associated with the vacuum in
other formalisms.\site{nonpert,burkardt93,burkardt97} The simplest
example of this approach is provided by a scalar field theory with
spontaneous symmetry breaking.  It is possible to shift the scalar
field and deal explicitly with a theory containing symmetry breaking
interactions.  In the simplest case
$\phi^3$ is the only relevant or marginal symmetry breaking
interaction, and one can simply tune this coupling to the value
corresponding to spontaneous rather than explicit symmetry breaking. 
Ken Wilson and I have also shown that in such simple cases one can use
coupling coherence to fix the strength of this interaction so that
tuning is not required.\site{coupcoh}

I will make an additional drastic assumption in these lectures, an
assumption that Ken Wilson does not believe will hold true.  I will
assume that all effective interactions we require are local in the
transverse direction.  If this is true, there are a finite number of
relevant and marginal operators, although each contains a function of
longitudinal momenta that must be determined by the renormalization
procedure.\footnote{These functions imply that there are effectively
an infinite number of relevant and marginal operators; however, their
dependence on fields and transverse momenta is extremely limited.}
There are {\it many more} relevant and marginal operators in the
renormalized light-front Hamiltonian than in ${\cal L}_{\rm QCD}$.
If transverse locality is violated, the situation is much worse than
this.

The presence of extra relevant and marginal operators that contain
functions tremendously complicates the renormalization problem, and
a common reaction to this problem is denial, which may persist for
years.  However, this situation may make possible tremendous
simplifications in the final nonperturbative problem.  For example,
few-body operators must produce confinement manifestly! 

Confinement cannot require particle creation and annihilation, flux
tubes, etc.  This is easily seen using a variational argument. 
Consider a color neutral quark-antiquark pair that are separated by a
distance $R$, which is slowly increased to infinity.  Moreover, to
see the simplest form of confinement assume that there are no light
quarks, so that the energy should increase indefinitely as they are
separated if the theory possesses confinement.  At each separation
the gluon components of the state adjust themselves to minimize the
energy.  But this means that the expectation value of the Hamiltonian
for a state with no gluons must exceed the energy of the state with
gluons, and therefore must diverge even more rapidly than the energy
of the true ground state.  This means that there must be a two-body
confining interaction in the Hamiltonian.  If the renormalization
procedure is unable to produce such confining two-body interactions,
the constituent picture will not arise.

Manifest gauge invariance and manifest rotational invariance require
all physical states to contain an arbitrarily large number of
constituents.  Gauge invariance is not manifest since we work in
light-cone gauge with the zero modes removed; and it is easy to see
that manifest rotational invariance requires an infinite number of
constituents.  Rotations about transverse axes are generated by
dynamic operators in interacting light-front field theories, operators
that create and annihilate particles.\site{dirac,chaone} No state
with a finite number of constituents rotates into itself or transforms
as a simple tensor under the action of such generators.  These
symmetries seem to imply that we can not obtain a constituent
approximation.

To cut this Gordian knot we employ cutoffs that violate gauge
invariance and covariance, symmetries which then must be restored by
effective interactions, and which need not be restored exactly.  A
familiar example of this approach is supplied by lattice gauge
theory, where rotational invariance is violated by the lattice.

\subsection{Simple Strategy}

We have recently employed a conceptually simple strategy to complete
bound state calculations.  The first step is to use a
perturbative similarity renormalization
group\site{glazek1,glazek2,Gl97a} and coupling
coherence\site{coupcoh,perryrg} to find the renormalized Hamiltonian
as an expansion in powers of the canonical coupling:
\begin{equation}
H^\Lambda = h_0 + g_\Lambda h_1^\Lambda + g_\Lambda^2 h_2^\Lambda +
\cdot \cdot \cdot
\end{equation}
We compute this series to a finite order, and to date have not required
any {\it ad hoc} assumptions to uniquely fix the Hamiltonian.  No
operators are added to the hamiltonian, so the hamiltonian is
completely determined by the underlying theory to this order. 

The second step is to employ bound state perturbation theory to solve
the eigenvalue problem.  The complete Hamiltonian contains every
interaction (although each is cut off) contained in the canonical
Hamiltonian, and many more.  We separate the Hamiltonian,
\begin{equation}
H^\Lambda=H_0^\Lambda+V^\Lambda \;,
\end{equation}
treating $H_0^\Lambda$ nonperturbatively and computing the effects of
$V^\Lambda$ in bound state perturbation theory. We must choose
$H_0^\Lambda$ and $\Lambda$ so that $H_0^\Lambda$ is {\it manageable} and to
minimize corrections from higher orders of $V^\Lambda$ within a
constituent approximation.

If a constituent approximation is valid {\it after} $\Lambda$ is
lowered to a critical value which must be determined, we may be
able to move all creation and annihilation operators to $V^\Lambda$. 
$H_0^\Lambda$ will include many-body interactions that do not change
particle number, and these interactions should be primarily
responsible for the constituent bound state structure.

There are several obvious flaws in this strategy. Chiral
symmetry-breaking operators, which must be included in the
Hamiltonian since we work entirely in the hidden symmetry phase of
the theory, do not appear at any finite order in the coupling.  These
operators must simply be added and tuned to fit spectra or fixed by a
non-perturbative renormalization
procedure.\site{nonpert,mustaki,Br94b} In addition, there are
perturbative errors in the strengths of all operators that do
appear.  We know from simple scaling arguments\site{wilson75} that
when $\Lambda$ is in the scaling regime:
\begin{itemize}
\item{small errors in relevant operators exponentiate in the output,}
\item{small errors in marginal operators produce comparable errors in
output,}
\item{small errors in irrelevant operators tend to decrease
exponentially in the output.}
\end{itemize}
This means that even if a relevant operator appears (\eg, a
constituent quark or gluon mass operator), we may need to tune its
strength to obtain reasonable results.  We have not had to do this,
but we have recently studied some of the effects of tuning a gluon
mass operator.\site{Sz97a}

To date this strategy has produced well-known results in
QED\site{brazil,BJ97a,BJ97c} through the Lamb shift, and
reasonable results for heavy quark bound states in
QCD.\site{brazil,Br96a,Br97a,Sz97a} The primary objective of
the remainder of these lectures is to review these results.  I first
use the Schwinger model as an illustration of a light-front bound
state calculation.  This model does not require renormalization, so
before turning to QED and QCD I discuss the renormalization procedure
that we have developed.

\section{LIGHT-FRONT SCHWINGER MODEL}

In this section I use the Schwinger model, massless QED in $1+1$
dimensions, to illustrate the basic strategy we employ {\it after} we
have computed the renormalized Hamiltonian.  No model in
$1+1$ dimensions illustrates the renormalization problems we
must solve before we can start to study QCD$_{3+1}$.  

The Schwinger model can be solved analytically.\site{schwing,
lowenstein} Charged particles are confined because the Coulomb
interaction is linear and there is only one physical particle, a
massive neutral scalar particle with no self-interactions.  The Fock
space content of the physical states depends crucially on the
coordinate system and gauge, and it is only in light-front coordinates
that a simple constituent picture emerges.\site{bergknoff}

The Schwinger model was first studied in Hamiltonian light-front field
theory by Bergknoff.\site{bergknoff} My description of the model
follows his closely, and I recommend his paper to the reader.
Bergknoff showed that the physical boson in the light-front massless
Schwinger model in light-cone gauge is a pure electron-positron
state.  This is an amazing result in a strong-coupling theory of
massless bare particles, and it illustrates how a constituent picture
may arise in QCD.  The electron-positron pair is confined by the
linear Coulomb potential. The light-front kinetic energy vanishes in
the massless limit, and the potential energy is minimized by a wave
function that is flat in momentum space, as one might expect since a
linear potential produces a state that is as localized as possible
(given kinematic constraints due to the finite velocity of light) in
position space.

In order to solve this theory I must first set up a large number of
details.  I recommend that for a first reading these details be
skimmed, because the general idea is more important than the detailed
manipulations.  The Lagrangian for the theory is
\begin{equation}
{\cal L} = \overline{\psi} \bigl( i \not{{\hskip-.08cm}\partial} -m
\bigr) \psi - e \overline{\psi} \gamma_\mu \psi A^\mu - {1 \over 4}
F_{\mu \nu} F^{\mu \nu} \;,
\end{equation}
where $F_{\mu \nu}$ is the electromagnetic field strength
tensor.  I have included an electron mass, $m$, which is taken to
zero later.  I choose light-cone gauge,
\begin{equation}
A^+=0 \;.
\end{equation}
In this gauge we avoid ghosts, so that the Fock space has a
positive norm.  This is absolutely essential if we want to apply
intuitive techniques from many-body quantum mechanics.

Many calculations are simplified by the use of a chiral representation
of the Dirac gamma matrices, so in this section I will use:
\begin{equation}
\gamma^0=\left(\begin{array}{cc}
                0 & -i \\ i & 0
                \end{array}\right)~~,~~~
                                 \gamma^1=\left(\begin{array}{cc}
                                                0 & i \\ i & 0
                                           \end{array}\right)\;,
\end{equation}
which leads to the light-front coordinate gamma matrices,
\begin{equation}
\gamma^+=\left(\begin{array}{cc}
                0 & 0 \\ 2 i & 0
                \end{array}\right)~~,~~~
                                 \gamma^-=\left(\begin{array}{cc}
                                                0 & -2 i \\ 0 & 0
                                           \end{array}\right)\;.
\end{equation}

In light-front coordinates the fermion field $\psi$
contains only one dynamical degree of freedom, rather than two.
To see this, first define the projection operators,
\begin{equation}
\Lambda_+={1 \over 2} \gamma^0 \gamma^+=\left(\begin{array}{cc}
                                               1 & 0 \\ 0 & 0
                                               \end{array}\right)~~~~~
 \Lambda_-={1 \over 2} \gamma^0 \gamma^-=\left(\begin{array}{cc}
                                                0 & 0 \\ 0 & 1
                                           \end{array}\right)\;.
\end{equation}
Using these operators split the fermion field into two
components,
\begin{equation}
\psi=\psi_+ + \psi_-=\Lambda_+ \psi + \Lambda_- \psi \;.
\end{equation}
The two-component Dirac equation in this gauge is
\begin{equation}
\biggl( {i \over 2} \gamma^+ \partial^- + {i \over 2} \gamma^-
\partial^+ -m - {e \over 2} \gamma^+ A^- \biggr) \psi = 0 \;;
\end{equation}
which can be split into two one-component equations,
\begin{equation}
i \partial^- \psit_+ = -i m \psit_- + e A^- \psit_+ \;,
\end{equation}
\begin{equation}
i \partial^+ \psit_- = i m \psit_+ \;.
\end{equation}
Here $\psit_\pm$ refers to the non-zero component of
$\psi_\pm$.

The equation for $\psi_+$ involves the light-front time derivative,
$\partial^-$; so $\psi_+$ is a dynamical degree of freedom that must
be quantized.  On the other hand, the equation for $\psi_-$ involves
only spatial derivatives, so $\psi_-$ is a constrained degree of
freedom that should be eliminated in favor of $\psi_+$.  Formally,
\begin{equation}
\psit_-={m \over \partial^+} \psit_+ \;.
\end{equation}
This equation is not well-defined until boundary conditions
are specified so that $\partial^+$ can be inverted.  I will eventually
define this operator in momentum space using a cutoff, but I want to
delay the introduction of a cutoff until a calculation requires it.

I have chosen the gauge so that $A^+=0$, and the equation for $A^-$
is
\begin{equation}
-{1 \over 4} \bigl(\partial^+\bigr)^2 A^- = e \psi_+^\dagger \psi_+ \;.
\end{equation}
$A^-$ is also a constrained degree of freedom, and we can
formally eliminate it,
\begin{equation}
A^-=-{4 e \over \bigl(\partial^+\bigr)^2} \psi_+^\dagger \psi_+ \;.
\end{equation}

We are now left with a single dynamical degree of freedom, $\psi_+$,
which we can quantize at $x^+=0$,
\begin{equation}
\bigl\{\psi_+(x^-),\psi_+^\dagger(y^-)\bigr\} = \Lambda_+
\delta(x^--y^-) \;.
\end{equation}
We can introduce free particle creation and annihilation
operators and expand the field operator at $x^+=0$,
\begin{equation}
\psit_+(x^-) = \int_{k^+ > 0} {dk^+ \over 4\pi} \Biggl[ b_k e^{-i k
\cdot x} + d_k^\dagger e^{i k \cdot x} \Biggr] \;,
\end{equation}
with,
\begin{equation}
\bigl\{b_k,b_p^\dagger\} = 4 \pi \delta(k^+-p^+) \;.
\end{equation}
In order to simplify notation, I will often write $k$ to
mean $k^+$.  If I need $k^-=m^2/k^+$, I will provide the superscript.

The next step is to formally specify the Hamiltonian.  I start with
the canonical Hamiltonian,
\begin{equation}
H = \int dx^- \Bigl( H_0 + V \Bigr) \;,
\end{equation}
\begin{equation}
H_0 = \psi_+^\dagger \Biggl({m^2 \over i\partial^+}\Biggr)
\psi_+ \;,
\end{equation}
\begin{equation}
V= -2 e^2 \psi_+^\dagger \psi_+ \Biggl({1 \over
\partial^+}\Biggr)^2 \psi_+^\dagger \psi_+ \;.
\end{equation}
To actually calculate:
\begin{itemize}
\item{replace $\psi_+$ with its expansion in terms of $b_k$ and $d_k$},
\item{normal-order},
\item{throw away constants},
\item{drop all operators that require $b_0$ and $d_0$}.
\end{itemize}
The free part of the Hamiltonian becomes
\begin{equation}
H_0=\int_{k>0} {dk \over 4\pi} \Biggl({m^2 \over k}\Biggr)
\bigl(b_k^\dagger b_k+d_k^\dagger d_k\bigr) \;.
\end{equation}

When $V$ is normal-ordered, we encounter new one-body operators,
\begin{equation}
H'_0={e^2 \over 2\pi} \int_{k>0} {dk \over 4\pi}
\Biggl[\int_{p>0} dp \biggl( {1 \over (k-p)^2} - {1 \over (k+p)^2}
\biggr)\Biggr] \bigl(b_k^\dagger b_k+d_k^\dagger d_k\bigr) \;.
\end{equation}
This operator contains a divergent momentum integral.  From
a mathematical point of view we have been sloppy and need to carefully
add boundary conditions and define how $\partial^+$ is inverted.
However, I want to apply physical intuition and even though no
physical photon has been exchanged to produce the initial interaction,
I will act as if a photon has been exchanged and everywhere an
`instantaneous photon exchange' occurs I will cut off the momentum.
In the above integral I insist,
\begin{equation}
|k-p|>\epsilon \;.
\end{equation}
Using this cutoff we find that
\begin{equation}
H'_0={e^2 \over \pi} \int{dk \over 4\pi} \biggl({1 \over
\epsilon} - {1 \over k} + {\cal O}(\epsilon) \biggr) \bigl(b_k^\dagger
b_k + d_k^\dagger d_k \bigr) \;.
\end{equation}

Comparing this result with the original free Hamiltonian, we see that
a divergent mass-like term appears; but it does not have the same
dispersion relation as the bare mass. Instead of depending on the
inverse momentum of the fermion, it depends on the inverse momentum
cutoff, which cannot appear in any physical result.  There is also a
finite shift in the bare mass, with the standard dispersion relation.

The normal-ordered interactions are
\begin{eqnarray}
V'= 2 e^2 \int {dk_1 \over 4\pi}\cdot\cdot\cdot{dk_4 \over 4\pi}
4\pi\delta(k_1+k_2-k_3-k_4) \nonumber \\
~~~~~~~~~~~ \Biggl\{ -{2 \over (k_1-k_3)^2} b_1^\dagger d_2^\dagger d_4
b_3 +
{2 \over (k_1+k_2)^2} b_1^\dagger d_2^\dagger d_3 b_4 \nonumber \\
~~~~~~~~~~~~~~-{1 \over (k_1-k_3)^2}
\bigl(b_1^\dagger b_2^\dagger b_3 b_4 +
d_1^\dagger d_2^\dagger d_3 d_4\bigr) +\cdot\cdot\cdot \Biggr\} \;.
\end{eqnarray}
I do not display the interactions that involve the creation
or annihilation of electron/positron pairs, which are important for
the study of multiple boson eigenstates.

The first term in $V'$ is the electron-positron interaction.
The longitudinal momentum cutoff I introduced above requires
$|k_1-k_3| > \epsilon$, so in position space we encounter a potential
which I will naively define with a Fourier transform that ignores the
fact that the momentum transfer cannot exceed the momentum of the
state,
\begin{eqnarray}
v(x^-) &=& 4 q_1 q_2 \int_{-\infty}^\infty {dk \over 4\pi}\; {1 \over
k^2}\;
\theta(|k|-\epsilon) \; {\rm exp}\bigl(-{i \over 2} k x^-\bigr)
\nonumber
\\ &=& {q_1 q_2 \over \pi} \; \Biggl[ {2 \over \epsilon} - {\pi \over
2} |x^-| + {\cal O}(\epsilon) \Biggr] \;.
\end{eqnarray}
This potential contains a linear Coulomb potential that we
expect in two dimensions, but it also contains a divergent constant
that is negative for unlike charges and positive for like charges.

In charge neutral states the infinite constant in $V'$ is {\it
exactly} canceled by the divergent `mass' term in $H'_0$. This
Hamiltonian assigns an infinite energy to states with net charge, and
a finite energy as $\epsilon \rightarrow 0$ to charge zero states.
This does not imply that charged particles are confined, but the
linear potential prevents charged particles from moving to arbitrarily
large separation except as charge neutral states.  The confinement
mechanism I propose for QCD in 3+1 dimensions shares many features
with this interaction.

I would also like to mention that even though the interaction between
charges is long-ranged, there are no van der Waals forces in 1+1
dimensions.  It is a simple geometrical calculation to show that all
long range forces between two neutral states cancel exactly.  This
does not happen in higher dimensions, and if we use long-range
two-body operators to implement confinement we must also find
many-body operators that cancel the strong long-range van der Waals
interactions.

Given the complete Hamiltonian in normal-ordered form we can study
bound states.  A powerful tool for the initial study of bound states
is the variational wave function.  In this case, we can begin with a
state that contains a single electron-positron pair,
\begin{equation}
|\Psi(P)\rangle = \int_0^P {dp \over 4\pi} \phi(p) b_p^\dagger
d_{P-p}^\dagger |0\rangle \;.
\end{equation}
The norm of this state is
\begin{equation}
\langle \Psi(P')|\Psi(P)\rangle = 4\pi P \delta (P'-P) \Biggl\{{1
\over P} \int_0^P {dp \over 4\pi} |\phi(p)|^2 \Biggr\}\;,
\end{equation}
where the factors outside the brackets provide a covariant
plane wave normalization for the center-of-mass motion of the bound
state, and the bracketed term should be set to one.

The expectation value of the one-body operators in the Hamiltonian is
\begin{equation}
\langle\Psi|H_0+H'_0|\Psi\rangle = {1 \over P} \int {dk
\over 4\pi} \Biggl[{m^2-e^2/\pi \over k}+ {m^2-e^2/\pi \over P-k}+
{2 e^2 \over \pi \epsilon} \Biggr] |\phi(k)|^2 \;,
\end{equation}
and the expectation value of the normal-ordered interactions
is
\begin{equation}
\langle\Psi|V'|\Psi\rangle = -{4 e^2 \over P} \int' {dk_1 \over
4\pi} {dk_2 \over 4\pi} \Bigl[{1 \over (k_1-k_2)^2}-{1 \over
P^2}\Biggr] \phi^*(k_1) \phi(k_2) \;,
\end{equation}
where I have dropped the overall plane wave norm. The prime
on the last integral indicates that the range of integration in which
$|k_1-k_2|<\epsilon$ must be removed.  By expanding the integrand
about $k_1=k_2$, one can easily confirm that the $1/\epsilon$
divergences cancel.

With $m=0$ the energy is minimized when
\begin{equation}
\phi(k)=\sqrt{4\pi} \;,
\end{equation}
and the invariant-mass is
\begin{equation}
M^2={e^2 \over \pi} \;.
\end{equation}

This type of simple analysis can be used to show that this
electron-positron state is actually the {\it exact ground} state of the
theory with momentum $P$, and that bound states do not interact with
one another.

The primary purpose of introducing the Schwinger model is to
illustrate that bound state center-of-mass motion is easily separated
from relative motion in light-front coordinates, and that standard
quantum mechanical techniques can be used to analyze the relative
motion of charged particles once the Hamiltonian is found.  It is
intriguing that even when the fermion is massless, the states are
constituent states in light-cone gauge and in light-front
coordinates.  This is not true in other gauges and coordinate systems.
The success of light-front field theory in 1+1 dimensions can
certainly be downplayed, but it should be emphasized that no other
method on the market is as powerful for bound state problems in 1+1
dimensions.

The most significant barriers to using light-front field theory to
solve low energy QCD are not encountered in 1+1 dimensions.  The
Schwinger model is super-renormalizable, so we completely avoid
serious ultraviolet divergences.  There are no transverse directions,
and we are not forced to introduce a cutoff that violates rotational
invariance, because there are no rotations.  Confinement results from
the Coulomb interaction, and chiral symmetry is not spontaneously
broken.  This simplicity disappears in realistic $3+1$-dimensional
calculations, which is one reason there are so few $3+1$-dimensional
light-front field theory calculations.

\section{LIGHT-FRONT RENORMALIZATION GROUP: SIMILARITY TRANSFORMATION
AND COUPLING COHERENCE}

As argued above, in $3+1$ dimensions we must introduce a cutoff on
energies, $\Lambda$, and we never perform explicit bound state
calculations with $\Lambda$ anywhere near its continuum limit. In
fact, we want to let $\Lambda$ become as small as possible.  In my
opinion, any strategy for solving light-front QCD that requires the
cutoff to explicitly approach infinity in the nonperturbative part of
the calculation is useless.  Therefore, we must set up and solve
\begin{equation}
P^-_\Lambda \mid \Psi_\Lambda(P) \rangle = {{\bf P}_\perp^2 + M^2
\over P^+} \mid \Psi_\Lambda(P) \rangle \;.
\end{equation}

Physical results, such as the mass, $M$, can not depend on the
arbitrary cutoff, $\Lambda$, {\it even} as $\Lambda$ approaches the
scale of interest.  This means that
$P^-_\Lambda$ and $\mid \Psi_\Lambda\rangle$ must depend on the cutoff
in such a way that $\langle \Psi_\Lambda \mid P^-_\Lambda \mid
\Psi_\Lambda \rangle$ does not.  Wilson based the derivation of his
renormalization group on this
observation,\site{wilson65,wilson70,wilson74,wilson75} and we use
Wilson's renormalization group to compute $P^-_\Lambda$.

It is difficult to even talk about how the Hamiltonian depends on the
cutoff without having a means of changing the cutoff.  If we can
change the cutoff, we can explicitly watch the Hamiltonian's cutoff
dependence change and fix its cutoff dependence by insisting that
this change satisfy certain requirements (\eg, that the limit in
which the cutoff is taken to infinity exists).  We introduce an
operator that changes the cutoff,
\begin{equation}
H(\Lambda_1) = T[H(\Lambda_0)] \;,
\end{equation}
where I assume that $\Lambda_1 < \Lambda_0$.  To simplify the
notation, I will let $H(\Lambda_l)=H_l$. To renormalize the
hamiltonian we study the properties of the transformation.

\begin{figure}
\begin{center}
\epsfig{file=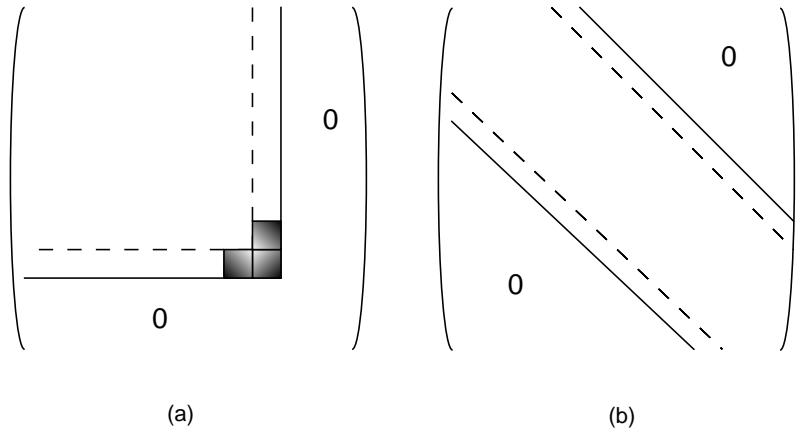,width=12cm}
\end{center} 
\caption{Two ways to run a cutoff on free energy.  In (a) a cutoff
on the magnitude of the energy is lowered from the solid to the
dashed lines, with problems resulting from the removed shaded
region. In (b) a cutoff on how far off diagonal matrix elements
appear is lowered from the dashed to the solid lines.}
\label{fig:fig3}
\end{figure}

Figure 3 displays two generic cutoffs that might be used.
Traditionally theorists have used cutoffs that remove high energy
states, as shown in Figure 3a.  This is the type of cutoff Wilson
employed in his initial work\site{wilson65} and I have studied its use
in light-front field theory.\site{perryrg} When a cutoff on energies
is reduced, all effects of couplings eliminated must be moved to
effective operators.  As we will see explicitly below, when these
effective operators are computed perturbatively they involve products
of matrix elements divided by energy denominators.  Expressions
closely resemble those encountered in standard perturbation theory,
with the second-order operator involving terms of the form
\begin{equation}
\delta V_{ij} \sim {\langle \phi_i \mid V \mid \phi_k \rangle \langle
\phi_k \mid V \mid \phi_j \rangle \over \epsilon_i-\epsilon_k}\;.
\end{equation}
This new effective interaction replaces missing couplings, so the
states $\phi_i$ and $\phi_j$ are retained and the state $\phi_k$ is
one of the states removed.  The problem comes from the shaded, lower
right-hand corner of the matrix, where the energy denominator
vanishes for states at the corner of the remaining matrix.  In this
corner we should use nearly degenerate perturbation theory rather than
perturbation theory, but to do this requires solving high energy
many-body problems nonperturbatively before solving the low energy
few-body problems.

An alternative cutoff, which does not actually remove any states and
which can be run by a similarity transformation\footnote{In deference
to the original work I will call this a similarity transformation even
though in all cases of interest to us it is a unitary
transformation.} is shown in Figure 3b.  This cutoff removes
couplings between states whose free energy differs by more than the
cutoff.  The advantage of this cutoff is that the effective operators
resulting from it contain energy denominators which are never smaller
than the cutoff, so that a perturbative approximation for the
effective Hamiltonian may work well.  I discuss a conceptually simple
similarity transformation that runs this cutoff below.

Given a cutoff and a transformation that runs the cutoff, we can
discuss how the Hamiltonian depends on the cutoff by studying how it
changes with the cutoff.  Our objective is to find a ``renormalized"
Hamiltonian, which should give the same results as an idealized
Hamiltonian in which the cutoff is infinite and which displays all of
the symmetries of the theory.  To state this in simple terms,
consider a sequence of Hamiltonians generated by repeated application
of the transformation,
\begin{equation}
H_0 \rightarrow H_1 \rightarrow H_2
\rightarrow \cdot \cdot \cdot \;.
\end{equation}
What we really want to do is fix the final value of $\Lambda$ at a
reasonable hadronic scale, and let $\Lambda_0$ approach infinity.  In
other words, we seek a Hamiltonian that survives an infinite number
of transformations.  In order to do this we need to understand what
happens when the transformation is applied to a broad class of
Hamiltonians.

Perturbative renormalization group analyses typically begin with the
identification of at least one fixed point, $H^*$.  A {\it fixed
point} is defined to be any Hamiltonian that satisfies the condition
\begin{equation}
H^*=T[H^*] \;.
\end{equation}
For perturbative renormalization groups the search for such
fixed points is relatively easy. If $H^*$ contains no interactions
(\ie, no terms with a product of more than two field operators), it
is called {\it Gaussian}. If $H^*$ has a massless eigenstate, it is
called {\it critical}. If a Gaussian fixed point has no mass term, it
is a {\it critical Gaussian} fixed point. If it has a mass term, this
mass must typically be infinite, in which case it is a {\it trivial
Gaussian} fixed point.  In lattice QCD the trajectory of renormalized
Hamiltonians stays `near' a critical Gaussian fixed point until the
lattice spacing becomes sufficiently large that a transition to
strong-coupling behavior occurs.  If $H^*$ contains only weak
interactions, it is called {\it near-Gaussian}, and we may be able to
use perturbation theory both to identify $H^*$ and to accurately
approximate `trajectories' of Hamiltonians near $H^*$. Of course, once
the trajectory leaves the region of $H^*$, it is generally necessary to
switch to a non-perturbative calculation of subsequent evolution.

Consider the immediate `neighborhood' of the fixed point, and assume
that the trajectory remains in this neighborhood.  This assumption
must be justified {\it a posteriori}, but if it is true we should write
\begin{equation}
H_l=H^*+\delta H_l \;,
\end{equation}
and consider the trajectory of small deviations $\delta H_l$.

As long as $\delta H_l$ is `sufficiently small,' we can use a
perturbative expansion in powers of $\delta H_l$, which leads us to
consider
\begin{equation}
\delta H_{l+1}= L \cdot \delta H_l + N[\delta H_l] \;.
\end{equation}
Here $L$ is the linear approximation of the full
transformation in the neighborhood of the fixed point, and $N[\delta
H_l]$ contains all contributions to $\delta H_{l+1}$ of $\order(\delta
H_l^2)$ and higher.

The object of the renormalization group calculation is to compute
trajectories and this requires a representation for $\delta H_l$.  The
problem of computing trajectories is one of the most common in
physics, and a convenient basis for the representation of $\delta H_l$
is provided by the eigenoperators of $L$, since $L$ dominates the
transformation near the fixed point.  These eigenoperators and their
eigenvalues are found by solving
\begin{equation}
L \cdot O_m=\lambda_m O_m \;.
\end{equation}
If $H^*$ is Gaussian or near-Gaussian it is usually
straightforward to find $L$, and its eigenoperators and eigenvalues.
This is not typically true if $H^*$ contains strong interactions, but
in QCD we hope to use a perturbative renormalization group in the
regime of asymptotic freedom, and the QCD ultraviolet fixed point is
apparently a critical Gaussian fixed point.  For light-front field
theory this linear transformation is a scaling of the transverse
coordinate, the eigenoperators are products of field operators and
transverse derivatives, and the eigenvalues are determined by the
transverse dimension of the operator.  All operators can include
both powers and inverse powers of longitudinal derivatives because
there is no longitudinal locality.\site{perryrg}

Using the eigenoperators of $L$ as a basis we can represent $\delta
H_l$,
\begin{equation}
\delta H_l = \sum_{m\in R} \mu_{m_l}O_m +\sum_{m\in M} g_{m_l}O_m+
\sum_{m\in I} w_{m_l}O_m \;.
\end{equation}
Here the operators $O_m$ with $m\in R$ are {\it relevant}
(\ie, $\lambda_m>1$), the operators $O_m$ with $m\in M$ are {\it
marginal} (\ie, $\lambda_m=1$), and the operators with $m\in I$ are
either {\it irrelevant} (\ie, $\lambda_m<1$) or become irrelevant
after many applications of the transformation.  The motivation behind
this nomenclature is made clear by considering repeated application of
$L$, which causes the relevant operators to grow exponentially, the
marginal operators to remain unchanged in strength, and the irrelevant
operators to decrease in magnitude exponentially.  There are technical
difficulties associated with the symmetry of $L$ and the completeness
of the eigenoperators that I ignore.\site{wegner}

For the purpose of illustration, let me assume that $\lambda_m=4$ for
all relevant operators, and $\lambda_m=1/4$ for all irrelevant
operators. The transformation can be represented by an infinite number
of coupled, nonlinear difference equations:
\begin{equation}
\mu_{m_{l+1}}=4 \mu_{m_l} + N_{\mu_m}[\mu_{m_l}, g_{m_l}, w_{m_l}] \;,
\end{equation}
\begin{equation}
g_{m_{l+1}}=g_{m_l} + N_{g_m}[\mu_{m_l}, g_{m_l}, w_{m_l}] \;,
\end{equation}
\begin{equation}
w_{m_{l+1}}={1 \over 4} w_{m_l} + N_{w_m}[\mu_{m_l}, g_{m_l},
w_{m_l}] \;.
\end{equation}
Sufficiently near a critical Gaussian fixed point, the
functions $N_{\mu_m}$, $N_{g_m}$, and $N_{w_m}$ should be adequately
approximated by an expansion in powers of $\mu_{m_l}$, $g_{m_l}$, and
$w_{m_l}$. The assumption that the Hamiltonian remains in the
neighborhood of the fixed point, so that all $\mu_{m_l}$, $g_{m_l}$,
and $w_{m_{l}}$ remain small, must be justified {\it a posteriori}.
Any precise definition of the neighborhood of the fixed point within
which all approximations are valid must also be provided {\it a
posteriori}.

Wilson has given a general discussion of how these equations can be
solved,\site{wilson75} but I will use coupling
coherence\site{coupcoh,perryrg} to fix the Hamiltonian.  This is
detailed below, so at this point I will merely state that coupling
coherence allows us to fix all couplings as functions of the
canonical couplings and masses in a theory.  The renormalization group
equations specify how all of the couplings run, and coupling
coherence uses this behavior to fix the strength of all of the
couplings.  But the first step is to develop a transformation.

\subsection{Similarity Transformation}

Stan G{\l}azek and Ken Wilson studied the problem of small energy
denominators which are apparent in Wilson's first complete
non-perturbative renormalization group calculations,\site{wilson70}
and realized that a similarity transformation which runs a different
form of cutoff (as discussed above) avoids this
problem.\site{glazek1,glazek2,Gl97a} Independently,
Wegner\site{wegner94} developed a similarity transformation which is
easier to use than that of G{\l}azek and Wilson.

In this section I want to give a simplified discussion of the
similarity transformation, using sharp cutoffs that must eventually be
replaced with smooth cutoffs which require a more complicated
formalism.

Suppose we have a Hamiltonian,
\begin{equation}
H^\lzero=H_0^\lzero+V^\lzero \;,
\end{equation}
where $H^\lzero_0$ is diagonal.  The cutoff $\lzero$
indicates that $\langle \phi_i|V^\lzero|\phi_j\rangle=0$ if $|E_{0
i}-E_{0 j}|>\lzero$. I should note that $\lzero$ is defined
differently in this section from later sections.  We want to use a
similarity transformation, which automatically leaves all eigenvalues
and other physical matrix elements invariant, that lowers this cutoff
to $\lone$.  This similarity transformation will constitute the first
step in a renormalization group transformation, with the second step
being a rescaling of energies that returns the cutoff to its original
numerical value.\footnote{The rescaling step is not essential, but it
avoids exponentials of the cutoff in the renormalization group
equations.}

The transformed Hamiltonian is
\begin{equation}
H^\lone=e^{i R}\bigl(H_0^\lzero+V^\lzero\bigr)e^{-i R} \;,
\end{equation}
where $R$ is a hermitian operator. If $H^\lzero$ is already
diagonal, $R=0$.  Thus, if $R$ has an expansion in powers of $V$, it
starts at first order and we can expand the exponents in powers of $R$
to find the perturbative approximation of the transformation.

We must adjust $R$ so that the matrix elements of $H^\lone$ vanish for
all states that satisfy $\lone<|E_{0 i}-E_{0 j}|<\lzero$.  We insist
that this happens to each order in perturbation theory.  Consider such
a matrix element,
\begin{eqnarray}
\langle \phi_i|H^\lone|\phi_j\rangle &=& \langle \phi_i|
 e^{i R}\bigl(H_0^\lzero+V^\lzero\bigr)e^{-i R} |\phi_j\rangle
\nonumber \\ &=& \langle \phi_i| (1+i R+\cdot\cdot\cdot)
(H_0^\lzero+V^\lzero\bigr) (1 -i R-\cdot\cdot\cdot) |\phi_j\rangle
\nonumber \\ &=& \langle \phi_i|H_0^\lzero |\phi_j\rangle  +
\langle \phi_i|V^\lzero+i\Bigl[R,H_0^\lzero\Bigr] |\phi_j\rangle +
\cdot\cdot\cdot \;.
\end{eqnarray}
The last line contains all terms that appear in first-order
perturbation theory.  Since $\langle \phi_i|H_0^\lzero |\phi_j\rangle
=0$ for these off-diagonal matrix elements, we can satisfy our new
constraint using
\begin{equation}
\langle \phi_i|R |\phi_j\rangle = {i \langle \phi_i|V^\lzero
|\phi_j\rangle \over E_{0 j}-E_{0 i}} \;+\;\order(V^2) \;.
\end{equation}
This fixes the matrix elements of $R$ when $\lone<|E_{0
i}-E_{0 j}|<\lzero$ to first order in $V^\lzero$.  I will assume that
the matrix elements of $R$ for $|E_{0i}-E_{0j}|<\lone$ are zero to
first order in $V$, and fix these matrix elements to second order
below.

Given $R$ we can compute the nonzero matrix elements of $H^\lone$.  To
second order in $V^\lzero$ these are
\begin{eqnarray}
H^\lone_{ab} &=& \langle \phi_a| H_0+V + i\Bigl[R,H_0\Bigr]
+i\Bigl[R,V \Bigr] -{1 \over 2} \Bigl\{R^2,H_0\Bigr\}
+ R H_0 R + \order(V^3) |\phi_b\rangle \nonumber \\
&=& \langle \phi_a|H_0+V + i\Bigl[R_2,H_0\Bigr]|\phi_b\rangle
\nonumber \\ &&
-{1 \over 2} \sum_k \Theta_{a k} \Theta_{k b} V_{ak} V_{kb} \Biggl[ {1
\over E_{0k}-E_{0a}}+{1 \over E_{0k}-E_{0b}} \Biggr] \nonumber \\
&&-\sum_k\Biggl[\Theta_{ak}\bigl(1-\Theta_{kb}\bigr) {V_{ak}
V_{kb} \over E_{0k}-E_{0a}} + \Theta_{kb}\bigl(1-\Theta_{ak}\bigr)
{V_{ak} V_{kb} \over E_{0k}-E_{0b}} \Biggr]+\order(V^3) \nonumber
\\
&=& \langle \phi_a|H_0+V+ i\Bigl[R_2,H_0\Bigr] |\phi_b\rangle
\nonumber \\&&+{1 \over 2} \sum_k \Theta_{a k} \Theta_{k b}
V_{ak} V_{kb} \Biggl[ {1 \over E_{0k}-E_{0a}}+{1 \over E_{0k}-E_{0b}}
\Biggr] \times \nonumber \\
&&~~~~~\Biggl[\theta(|E_{0a}-E_{0k}|-|E_{0b}-E_{0k}|) -
\theta(|E_{0b}-E_{0k}|- |E_{0a}-E_{0k}|)\Biggr] \nonumber \\
&&-\sum_k V_{ak} V_{kb} \biggl[ \Theta_{ak} {
\theta(|E_{0a}-E_{0k}|-|E_{0b}-E_{0k}|) \over E_{0k} - E_{0a} } +
\nonumber \\
&&~~~~~~~~~~~~~~~~~~~~~~~~ \Theta_{kb} { \theta(|E_{0b}-E_{0k}|-
|E_{0a}-E_{0k}|) \over E_{0k} - E_{0b} } \Biggr] \;.
\end{eqnarray}
I have dropped the $\lzero$ superscript on the right-hand
side of this equation and used subscripts to indicate matrix
elements. The operator $\Theta_{ij}$ is one if
$\lone<|E_{0i}-E_{0j}|<\lzero$ and zero otherwise.  It should also be
noted that $V_{ij}$ is zero if $|E_{0i}-E_{0j}|>\lzero$.  All energy
denominators involve energy differences that are at least as large as
$\lone$, and this feature persists to higher orders in perturbation
theory; which is the main motivation for choosing this transformation.
$R_2$ is second-order in $V$, and we are still free to choose its
matrix elements; however, we must be careful not to introduce small
energy denominators when choosing $R_2$.

The matrix element $\langle \phi_a| i\Bigl[R_2, H_0 \Bigr] |\phi_b
\rangle = i (E_{0b}-E_{0a}) \langle \phi_a| R_2 | \phi_b \rangle $
must be specified.  I will choose this matrix element to cancel the
first sum in the final right-hand side of Eq. (55).   This choice
leads to the same result one obtains by integrating a differential
transformation that runs a step function cutoff. To cancel the first
sum in the final right-hand side of Eq. (55) requires
\begin{eqnarray}
\langle \phi_a| R_2 | \phi_b \rangle &&= {i \over 2} \sum_k \Theta_{ak}
\Theta_{kb} V_{ak} V_{kb} \Biggl[ {1 \over E_{0k}-E_{0a}}+{1 \over
E_{0k}-E_{0b}} \Biggr] \times \nonumber \\
&&\Biggl[\theta(|E_{0a}-E_{0k}|-|E_{0b}-E_{0k}|) -
\theta(|E_{0b}-E_{0k}|- |E_{0a}-E_{0k}|)\Biggr] \;.
\end{eqnarray}
No small energy denominator appears in $R_2$ because it is
being used to cancel a term that involves a large energy difference. 
If we tried to use $R_2$ to cancel the remaining sum also, we would
find that it includes matrix elements that diverge as $E_{0a}-E_{0b}$
goes to zero, and this is not allowed.

The non-vanishing matrix elements of $H^{\Lambda_1}$ are now
completely determined to $\order(V^2)$,
\begin{eqnarray}
H^\lone_{ab} &=&
\langle \phi_a|H_0+V^\lzero|\phi_b\rangle
\nonumber \\&& - \sum_k V_{ak} V_{kb} \biggl[ \Theta_{ak} {
\theta(|E_{0a}-E_{0k}|-|E_{0b}-E_{0k}|) \over E_{0k} - E_{0a} } +
\nonumber \\
&&~~~~~~~~~~~~~~~~~~~~~~~~ \Theta_{kb} { \theta(|E_{0b}-E_{0k}|-
|E_{0a}-E_{0k}|) \over E_{0k} - E_{0b} } \Biggr]
 \;.
\end{eqnarray}
Let me again mention that $V^\lzero_{ij}$ is zero if
$|E_{0i}- E_{0j}|>\lzero$, so there are implicit cutoffs that result
from previous transformations.

As a final word of caution, I should mention that the use of step
functions produces long-range pathologies in the interactions that
lead to infrared divergences in gauge theories.  We must replace the
step functions with smooth functions to avoid this problem.  This
problem will not show up in any calculations detailed in these
lectures, but it does affect higher order calculations in
QED\site{BJ97a,BJ97c} and QCD.\site{BA97a}

\subsection{Light-Front Renormalization Group}

In this section I use the similarity transformation to form a
perturbative light-front renormalization group for scalar field
theory.  If we want to stay as close as possible to the canonical
construction of field theories, we can:
\begin{itemize}
\item Write a set of `allowed' operators using powers of
derivatives and field operators.
\item Introduce `free' particle creation and annihilation
operators, and expand all field operators in this basis.
\item Introduce cutoffs on the Fock space transition operators.
\end{itemize}

Instead of following this program I will skip to the final step, and
simply write a Hamiltonian to initiate the analysis.
\begin{eqnarray}
H & = &\qquad  \int \dqt_1 \; \dqt_2 \; (16 \pi^3)
\delta^3(q_1-q_2) \; u_2(-q_1,q_2)
\;a^\dagger(q_1) a(q_2) \nonumber \\
&&+{1 \over 6} \int \dqt_1\; \dqt_2\; \dqt_3\; \dqt_4 \; (16 \pi^3)
\delta^3(q_1+q_2+q_3-q_4) \nonumber \Theta(q_4^--q_3^--q_2^--q_1^-) \\
&&\qquad\qquad\qquad\qquad\qquad
u_4(-q_1,-q_2,-q_3,q_4)\; a^\dagger(q_1) a^\dagger(q_2)
a^\dagger(q_3)
a(q_4)  \nonumber \\
&&+{1 \over 4} \int \dqt_1\; \dqt_2\; \dqt_3\; \dqt_4 \; (16 \pi^3)
\delta^3(q_1+q_2-q_3-q_4) \Theta(q_4^-+q_3^--q_2^--q_1^-) \nonumber \\
&&\qquad\qquad\qquad\qquad\qquad u_4(-q_1,-q_2,q_3,q_4)\; a^\dagger(q_1)
a^\dagger(q_2) a(q_3) a(q_4)  \nonumber \\
&&+{1 \over 6} \int \dqt_1\; \dqt_2\; \dqt_3\; \dqt_4 \; (16 \pi^3)
\delta^3(q_1-q_2-q_3-q_4) \Theta(q_4^-+q_3^-+q_2^--q_1^-) \nonumber \\
&&\qquad\qquad\qquad\qquad\qquad u_4(-q_1,q_2,q_3,q_4)\; a^\dagger(q_1)
a(q_2) a(q_3) a(q_4)  \nonumber \\
&&+ \qquad {\cal O}(\phi^6) \;,
\end{eqnarray}
where,
\begin{equation}
\dqt={dq^+ d^2q_\perp \over 16\pi^3 q^+} \;,
\end{equation}
\begin{equation}
q_i^-={q_{i\perp}^2 \over q_i^+} \;,
\end{equation}
and,
\begin{equation}
\Theta(Q^-)=\theta\Biggl({\Lambda^2 \over P^+}-|Q^-| \Biggr) \;.
\end{equation}
I assume that no operators
break the discrete $\phi \rightarrow -\phi$ symmetry.  The functions
$u_2$ and $u_4$ are not yet determined.  If we assume locality in the
transverse direction, these functions can be expanded in powers of
their transverse momentum arguments.  Note that to specify the cutoff
both transverse and longitudinal momentum scales are required, and in
this case the longitudinal momentum scale is independent of the
particular state being studied.  Note also that $P^+$ breaks
longitudinal boost invariance and that a change in $P^+$ can be
compensated by a change in $\Lambda^2$.  This may have important
consequences, because the Hamiltonian should be a fixed point with
respect to changes in the cutoff's longitudinal momentum scale, since
this scale invariance is protected by Lorentz
covariance.\site{perryrg}

I have specified the similarity transformation in terms of matrix
elements, and will work directly with matrix elements, which are easily
computed in the free particle Fock space basis.  In order to study the
renormalization group transformation I will assume that the
Hamiltonian includes only the interactions shown above.  A single
transformation will produce a Hamiltonian containing products of
arbitrarily many creation and annihilation operators, but it is not
necessary to understand the transformation in full detail.

I will {\it define} the full renormalization group transformation as:
(i) a similarity transformation that lowers the cutoff in $\Theta$,
(ii) a rescaling of all transverse momenta that returns the cutoff to
its original numerical value, (iii) a rescaling of the creation and
annihilation operators by a constant factor $\zeta$, and (iv) an
overall constant rescaling of the Hamiltonian to absorb a
multiplicative factor that results from the fact that it has the
dimension of transverse momentum squared.  These rescaling operations
are introduced so that it may be possible to find a fixed point
Hamiltonian that contains interactions.\site{wilson75}

To find the critical Gaussian fixed point we need to study the
linearized approximation of the full transformation, as discussed
above.  In general the linearized approximation can be extremely
complicated, but near a critical Gaussian fixed point it is
particularly simple in light-front field theory with zero modes
removed, because tadpoles are excluded. We have already seen that the
similarity transformation does not produce any first order change in
the Hamiltonian (see Eq. (57)), so the first order change is
determined entirely by the rescaling operation.  If we let
\begin{equation}
\Lambda_1 = \eta \Lambda_0 \;,
\end{equation}
\begin{equation}
{\bf p}_{i\perp} = \eta {\bf p}_{i\perp}' \;,
\end{equation}
and,
\begin{equation}
a_p = \zeta a_{p'} \;,\;\;\;a_p^\dagger = \zeta a_{p'}^\dagger \;,
\end{equation}
to first order the transformed Hamiltonian is
\begin{eqnarray}
H & = &\qquad  \zeta^2 \int \dqt_1 \; \dqt_2 \; (16 \pi^3)
\delta^3(q_1-q_2) \; u_2(q_i^+, \eta q_i^\perp)
\;a^\dagger(q_1) a(q_2) \nonumber \\
&&+{1 \over 6} \eta^4 \zeta^4 \int \dqt_1\; \dqt_2\; \dqt_3\; \dqt_4 \; (16
\pi^3)
\delta^3(q_1+q_2+q_3-q_4) \nonumber \Theta(q_4^--q_3^--q_2^--q_1^-) \\
&&\qquad\qquad\qquad\qquad\qquad
u_4(q_i^+, \eta q_i^\perp)\; a^\dagger(q_1) a^\dagger(q_2)
a^\dagger(q_3)
a(q_4)  \nonumber \\
&&+{1 \over 4} \eta^4 \zeta^4 \int \dqt_1\; \dqt_2\; \dqt_3\; \dqt_4 \; (16
\pi^3)
\delta^3(q_1+q_2-q_3-q_4) \Theta(q_4^-+q_3^--q_2^--q_1^-) \nonumber \\
&&\qquad\qquad\qquad\qquad\qquad u_4(q_i^+, \eta q_i^\perp)\; a^\dagger(q_1)
a^\dagger(q_2) a(q_3) a(q_4)  \nonumber \\
&&+{1 \over 6} \eta^4 \zeta^4 \int \dqt_1\; \dqt_2\; \dqt_3\; \dqt_4 \; (16
\pi^3)
\delta^3(q_1-q_2-q_3-q_4) \Theta(q_4^-+q_3^-+q_2^--q_1^-) \nonumber \\
&&\qquad\qquad\qquad\qquad\qquad u_4(q_i^+, \eta q_i^\perp)\; a^\dagger(q_1)
a(q_2) a(q_3) a(q_4)  \nonumber \\
&&+ \qquad \order(\phi^6) \;,
\end{eqnarray}
where I have simplified my notation for the arguments
appearing in the functions $u_i$. An overall factor of $\eta^2$ that
results from the engineering dimension of the Hamiltonian has been
removed.  The Gaussian fixed point is found by insisting that the
first term remains constant, which requires
\begin{equation}
\zeta^2 u_2^*(q^+, \eta {\bf q}_\perp)=u_2^*(q^+,{\bf q}_\perp) \;.
\end{equation}
I have used the fact that $u_2$ actually depends on only one
momentum other than the cutoff.

The solution to this equation is a monomial in ${\bf q}_\perp$ which
depends on $\zeta$,
\begin{equation}
u_2^*(q^+,{\bf q}_\perp)=f(q^+)\bigl({\bf q}_\perp \bigr)^n \;,\;\;\;
\zeta=\Biggl({1 \over \eta}\Biggr)^{(n/2)} \;.
\end{equation}
The solution depends on our choice of $n$ and to obtain the
appropriate free particle dispersion relation we need to choose $n=2$,
so that
\begin{equation}
u_2^*(q^+,{\bf q}_\perp)=f(q^+) {\bf q}_\perp^2 \;\;,\;\;\;
\zeta={1 \over \eta} \;.
\end{equation}
$f(q^+)$ is allowed because the cutoff scale $P^+$ allows us
to form the dimensionless variable $q^+/P^+$, which can enter the
one-body operator.  We will see that this happens in QED and QCD. Note
that the constant in front of each four-point interaction becomes one,
so that their scaling behavior is determined entirely by $u_4$. If we
insist on transverse locality (which may be violated because we remove
zero modes), we can expand $u_4$ in powers of its transverse momentum
arguments, and discover powers of $\eta q_i^\perp$ in the transformed
Hamiltonian. Since we are lowering the cutoff, $\eta<1$, and each
power of transverse momentum will be suppressed by this factor. This
means increasing powers of transverse momentum are increasingly {\it
irrelevant}.

I will not go through a complete derivation of the eigenoperators of
the linearized approximation to the renormalization group
transformation about the critical Gaussian fixed point, but the
derivation is simple.\site{perryrg} Increasing powers of transverse
derivatives and increasing powers of creation and annihilation
operators lead to increasingly irrelevant operators.  The irrelevant
operators are called `non-renormalizable' in old-fashioned Feynman
perturbation theory.  Their magnitude decreases at an exponential rate
as the cutoff is lowered, which means that they {\it increase} at an
exponential rate as the cutoff is raised and produce increasingly
large divergences if we try to follow their evolution perturbatively
in this exponentially unstable direction.

The only relevant operator is the mass operator,
\begin{equation}
\mu^2 \int \dqt_1 \; \dqt_2 \; (16 \pi^3) \delta^3(q_1-q_2) \;
\;a^\dagger(q_1) a(q_2) \;,
\end{equation}
while the fixed point Hamiltonian is marginal (of course),
and the operator in which $u_4=\lambda$ (a constant) is marginal.  A
$\phi^3$ operator would also be relevant.

The next logical step in a renormalization group analysis is to study
the transformation to second order in the interaction, keeping the
second-order corrections from the similarity transformation.  I will
compute the correction to $u_2$ to this order and refer the interested
reader to Ref. \cite{perryrg} for more complicated examples.

The matrix element of the one-body operator between single
particle states is
\begin{equation}
\langle p'|h|p\rangle = \langle 0|\;a(p')\; h\;
a^\dagger(p)\;|0\rangle = (16 \pi^3) \delta^3(p'-p)\;u_2(-p,p) \;.
\end{equation}
Thus, we easily determine $u_2$ from the matrix element.  It
is easy to compute matrix elements between other states.  We computed
the matrix elements of the effective Hamiltonian generated by the
similarity transformation when the cutoff is lowered in Eq. (57), and
now we want to compute the second-order term generated by the
four-point interactions above.  There are additional corrections to
$u_2$ at second order in the interaction if $u_6$, {\it etc.} are
nonzero.

Before rescaling we find that the transformed Hamiltonian contains
\begin{eqnarray}
&&(16 \pi^3)\delta^3(p'-p)\;\delta v_2(-p,p)=
\nonumber \\ &&~~~~~
{1 \over 3!} \int \dkt_1 \dkt_2 \dkt_3 \; \theta\bigl
(\Lambda_0 -| p^- - k_1^- -k_2^- -k_3^-|\bigr) \nonumber \\
&&~~~~~
\theta\bigl( | p^- - k_1^- -k_2^- -k_3^-| - \eta \Lambda_0\bigr) ~
{\langle p'| V|k_1,k_2,k_3\rangle \langle k_1,k_2,k_3|V|p\rangle \over
p^--k_1^--k_2^--k_3^-} \;,
\end{eqnarray}
where $p^-={\bf p}_\perp^2/p^+$, {\it etc.} One can readily
verify that
\begin{equation}
\langle p'|V|k_1,k_2,k_3\rangle = (16 \pi^3)
\delta^3(p'-k_1-k_2-k_3) \; \Theta\bigl(p^--k_1^--k_2^--k_3^-\bigr)\;
u_4(-p',k_1,k_2,k_3) \;,
\end{equation}
\begin{equation}
\langle k_1,k_2,k_3|V|p\rangle= (16 \pi^3) \delta^3(p-k_1-k_2-k_3) \;
\Theta\bigl(p^--k_1^--k_2^--k_3^-\bigr)\; u_4(-k_1,-k_2,-k_3,p) \;.
\end{equation}
This leads to the result,
\begin{eqnarray}
\delta v_2(-p,p)&=&{1 \over 3!} \int \dkt_1 \dkt_2 \dkt_3 \;
(16 \pi^3)\delta^3(p-k_1-k_2-k_3) \nonumber \\
&&\theta\bigl(\Lambda_0 -| p^- - k_1^- -k_2^- -k_3^-|\bigr)~
\theta\bigl( | p^- - k_1^- -k_2^- -k_3^-| - \eta \Lambda_0\bigr)
\nonumber \\ &&~~~~~~~
{u_4(-p,k_1,k_2,k_3)\; u_4(-k_1,-k_2,-k_3,p) \over
p^--k_1^--k_2^--k_3^-} \;.
\end{eqnarray}
To obtain $\delta u_2$ from $\delta v_2$ we must rescale the momenta,
the fields, and the Hamiltonian.  The final result is
\begin{eqnarray}
\delta u_2(-p,p)&=&{1 \over 3!} \int \dkt_1 \dkt_2 \dkt_3 \;
(16 \pi^3)\delta^3(p-k_1-k_2-k_3) \nonumber \\
&&\theta\Biggl({\Lambda_0 \over \eta} -| p^- - k_1^- -k_2^-
-k_3^-|\Biggr)~
\theta\bigl( | p^- - k_1^- -k_2^- -k_3^-| - \Lambda_0\bigr)
\nonumber \\ &&~~~~~~~
{u_4(p',k_1',k_2',k_3')\; u_4(-k_1',-k_2',-k_3',p') \over
p^--k_1^--k_2^--k_3^-} \;,
\end{eqnarray}
where $p^{+'}=p^+$, $p^{\perp'}=\eta p^\perp$,
$k_i^{+'}=k_i^+$, and $k_i^{\perp'}=\eta k_i^\perp$.

\subsection{Coupling Coherence}

The basic mathematical idea behind coupling coherence was
first formulated by Oehme, Sibold, and Zimmerman.\site{coupcoh2}
They were interested in field theories where many couplings appear,
such as the standard model, and wanted to find some means of reducing
the number of couplings.  Wilson and I developed the ideas
independently in an attempt to deal with the functions that appear in
marginal and relevant light-front operators.\site{coupcoh}

The puzzle is how to reconcile our knowledge from covariant
formulations of QCD that only one running coupling constant
characterizes the renormalized theory with the appearance of new
counterterms and functions required by the light-front formulation.
What happens in perturbation theory when there are effectively an
infinite number of relevant and marginal operators?  In particular,
does the solution of the perturbative renormalization group equations
require an infinite number of independent counterterms (\ie,
independent functions of the cutoff)? Coupling coherence provides the
conditions under which a finite number of running variables
determines the renormalization group trajectory of the renormalized
Hamiltonian. To leading nontrivial orders these conditions are
satisfied by the counterterms introduced to restore Lorentz
covariance in scalar field theory and gauge invariance in light-front
gauge theories. In fact, the conditions can be used to determine all
counterterms in the Hamiltonian, including relevant and marginal
operators that contain functions of longitudinal momentum fractions;
and with no direct reference to Lorentz covariance, this symmetry is
restored to observables by the resultant counterterms in scalar field
theory.\site{perryrg}

A coupling-coherent Hamiltonian is analogous to a fixed point
Hamiltonian, but instead of reproducing itself exactly it reproduces
itself in form with a limited number of independent running couplings. 
If $g_\Lambda$ is the only independent coupling in a theory, in a
coupling-coherent Hamiltonian {\it all other couplings are invariant
functions of $g_\Lambda$, $f_i(g_\Lambda)$}.  The couplings
$f_i(g_\Lambda)$ depend on the cutoff only through their dependence
on the running coupling $g_\Lambda$, and in general we demand
$f_i(0)=0$.  This boundary condition on the dependent couplings is
motivated in our calculations by the fact that it is the combination
of the cutoff and the interactions that force us to add the
counterterms we seek, so the counterterms should vanish when the
interactions are turned off.

Let me start with a simple example in which there is a finite
number of relevant and marginal operators \abinit, and use coupling
coherence to discover when only one or two of these may independently
run with the cutoff.  In general such conditions are met only when an
underlying symmetry exists.

Consider a theory in which two scalar fields interact,
\begin{equation}
V(\phi)={\lambda_1 \over 4!}\phi_1^4+{\lambda_2 \over 4!}\phi_2^4+
{\lambda_3 \over 4!}\phi_1^2 \phi_2^2 \;.
\end{equation}
Under what conditions will there be fewer than three
independent running coupling constants?  We can use a simple cutoff on
Euclidean momenta,
$q^2<\Lambda^2$.  Letting $t=\ln(\Lambda/\Lambda_0)$, the
Gell-Mann--Low equations are
\begin{equation}
{\partial \lambda_1 \over \partial t} = 3 \zeta \lambda_1^2 + {1
\over 12} \zeta \lambda_3^2 + \order(2\;{\rm loop}) \;,
\end{equation}
\begin{equation}
{\partial \lambda_2 \over \partial t} = 3 \zeta \lambda_2^2 + {1
\over 12} \zeta \lambda_3^2 + \order(2\;{\rm loop}) \;,
\end{equation}
\begin{equation}
{\partial \lambda_3 \over \partial t} = {2 \over 3}
\zeta \lambda_3^2 +  \zeta \lambda_1 \lambda_3+\zeta \lambda_2
\lambda_3 + \order(2\;{\rm loop}) \;;
\end{equation}
where $\zeta=\hbar/(16\pi^2)$.  It is not important at this
point to understand how these equations are derived.

First suppose that $\lambda_1$ and $\lambda_2$ run separately, and ask
whether it is possible to find $\lambda_3(\lambda_1,\lambda_2)$ that
solves Eq. (79).  To one-loop order this leads to
\begin{equation}
\Bigl(3 \lambda_1^2 + {1 \over 12} \lambda_3^2 \Bigr) \;
{\partial \lambda_3 \over \partial \lambda_1} \;+\;
\Bigl(3 \lambda_2^2 + {1 \over 12} \lambda_3^2 \Bigr) \;
{\partial \lambda_3 \over \partial \lambda_2} =
{2 \over 3} \lambda_3^2 +  \lambda_1 \lambda_3+
\lambda_2 \lambda_3 \;.
\end{equation}
If $\lambda_1$ and $\lambda_2$ are independent, we can
equate powers of these variables on each side of Eq. (80).  If we
allow the expansion of $\lambda_3$ to begin with a constant, we find a
solution to Eq. (80) in which all powers of $\lambda_1$ and
$\lambda_2$ appear. In this case a constant appears on the
right-hand-sides of Eqs. (77) and (78), and there will be no Gaussian
fixed points for
$\lambda_1$ and $\lambda_2$. We are generally not interested in the
possibility that a counterterm does not vanish when the canonical
coupling vanishes, so we will simply discard this solution both here
and below.  We are interested in the conditions under which one
variable ceases to be independent, and the appearance of such an
arbitrary constant indicates that the variable remains independent
even though its dependence on the cutoff is being reparameterized in
terms of other variables.

If we do not allow a constant in the solution, we find that
$\lambda_3=\alpha \lambda_1+\beta \lambda_2+\order(\lambda^2)$. When
we insert this in Eq. (80) and equate powers on each side, we obtain
three coupled equations for $\alpha$ and $\beta$.  These equations
have no solution other than $\alpha=0$ and $\beta=0$, so we conclude
that if $\lambda_1$ and $\lambda_2$ are independent functions of $t$,
$\lambda_3$ will also be an independent function of $t$ unless the two
fields decouple.

Assume that there is only one independent variable, $\lt=\lambda_1$,
so that $\lambda_2$ and $\lambda_3$ are functions of $\lt$.  In this
case we obtain two coupled equations,
\begin{equation}
\Bigl(3 \lt^2+{1 \over 12} \lambda_3^2 \Bigr) \; {\partial \lambda_2
\over \partial \lt} = 3 \lambda_2^2+{1 \over 12} \lambda_3^2 \;,
\end{equation}
\begin{equation}
\Bigl(3 \lt^2+{1 \over 12} \lambda_3^2 \Bigr) \; {\partial \lambda_3
\over \partial \lt} = {2 \over 3} \lambda_3^2 + \lt \lambda_3+
\lambda_2 \lambda_3 \;.
\end{equation}
If we again exclude a constant term in the expansions of
$\lambda_2$ and $\lambda_3$ we find that the only non-trivial
solutions to leading order are $\lambda_2=\lt$, and either $\lambda_3
= 2 \lt$ or $\lambda_3=6 \lt$.  If $\lambda_3=2\lt$,
\begin{equation}
V(\phi)={\lt \over 4!}\;\bigl(\phi_1^2+\phi_2^2 \bigr)^2 \;,
\end{equation}
and we find the $O(2)$ symmetric theory.  If
$\lambda_3=6\lt$,
\begin{equation}
V(\phi)={\lt \over 2 \cdot 4!}\;\Bigl[\bigl(\phi_1+\phi_2\bigr)^4+
\bigl(\phi_1-\phi_2\bigr)^4\Bigr] \;,
\end{equation}
and we find two decoupled scalar fields.  Therefore,
$\lambda_2$ and $\lambda_3$ do not run independently with the cutoff
if there is a symmetry that relates their strength to $\lambda_1$.

The condition that a limited number of variables run with the cutoff
does not only reveal symmetries broken by the regulator, it may also
be used to uncover symmetries that are broken by the vacuum.  I will
not go through the details, but it is straightforward to show that in
a scalar theory with a $\phi^3$ coupling, this coupling can be fixed
as a function of the $\phi^2$ and $\phi^4$ couplings only if the
symmetry is spontaneously broken rather than explicitly
broken.\site{coupcoh}

This example is of some interest in light-front field theory,
because it is difficult to reconcile vacuum symmetry breaking with the
requirements that we work with a trivial vacuum and drop zero-modes
in practical non-perturbative Hamiltonian calculations.  Of course,
the only way that we can build vacuum symmetry breaking into the
theory without including a nontrivial vacuum as part of the state
vectors is to include symmetry breaking interactions in the
Hamiltonian and work in the hidden symmetry phase.  The problem then
becomes one of finding all necessary operators without sacrificing
predictive power.  The renormalization group specifies what operators
are relevant, marginal, and irrelevant; and coupling coherence
provides one way to fix the strength of the symmetry-breaking
interactions in terms of the symmetry-preserving interactions.  This
does not solve the problem of how to treat the vacuum in light-front
QCD by any means, because we have only studied perturbation theory;
but this result is encouraging and should motivate further
investigation.

For the QED and QCD calculations discussed below, I need to compute
the hamiltonian to second order, while the canonical coupling runs at
third order.  To determine the generic solution to this problem, I
present an oversimplified analysis in which there are three
coupled renormalization group equations for the independent marginal
coupling ($g$), in addition to dependent relevant ($\mu$) and
irrelevant ($w$) couplings.
\begin{equation}
\mu_{m_{l+1}}=4 \mu_{m_l} + c_\mu g_l^2 + \order(g_l^3) \;,
\end{equation}
\begin{equation}
g_{m_{l+1}}=g_{m_l} + c_g g_l^3 + \order(g_l^4) \;,
\end{equation}
\begin{equation}
w_{m_{l+1}}={1 \over 4} w_{m_l} + c_w g_l^2 + \order(g_l^3) \;.
\end{equation}

I assume that $\mu_l = \zeta g_l^2 + \order(g_l^3)$ and $w_l=\eta
g_l^2 + \order(g_l^3)$, which satisfy the conditions of coupling
coherence.  Substituting into the renormalization group equations
yields (dropping all terms of $\order(g_l^3)$),
\begin{equation}
\zeta g_{l+1}^2 = \zeta g_l^2 = 4 \zeta g_l^2 + c_\mu g_l^2 \;,
\end{equation}
\begin{equation}
\eta g_{l+1}^2 = \eta g_l^2 = {1 \over 4} \eta g_l^2 + c_w g_l^2 \;.
\end{equation}
The solutions are $\zeta=-{1 \over 3} c_\mu$ and $\eta = {4
\over 3} c_w$.  

These are {\it exactly} the coefficients in a Taylor
series expansion for $\mu(g)$ and $w(g)$ that reproduce themselves.
This observation suggests an alternative way to find the coupling
coherent Hamiltonian without explicitly setting up the
renormalization group equations. Although this method is less
general, we only need to find what operators must be added to the
Hamiltonian so that at $\order(g^2)$ it reproduces itself, with the
only change being the change in the specific value of the cutoff. 
Coupling coherence allows us to substitute the running coupling in
this solution, but it is not until third order that we would
explicitly see the coupling run. This is how I will fix the
QED and QCD hamiltonians to second order.

\section{QED and QCD Hamiltonians}

In order to derive the renormalized QED and QCD Hamiltonians I must
first list the canonical Hamiltonians.  I follow the conventions of
Brodsky and Lepage.\site{brodsky} There is no need to be overly
rigorous, because coupling coherence will fix any perturbative
errors.  I recommend the papers of Zhang and Harindranath\site{Zh93a}
for a more detailed discussion from a different point of view.  The
reader who is not yet concerned with details can skip this section.

I will use $A^+=0$ gauge, and I drop zero modes. I use the Bjorken
and Drell conventions for gamma matrices.  The gamma matrices are
\begin{equation}
\gamma^0=\left(\begin{array}{cc}
                1 & 0 \\ 0 & -1
                \end{array}\right)~,~~~~
                                 \gamma^k=\left(\begin{array}{cc}
                                         0 & \sigma_k \\ -\sigma_k & 0
                                           \end{array}\right)\;,
\end{equation}
where $\sigma_k$ are the Pauli matrices.  This leads to
\begin{equation}
\gamma^+=\left(\begin{array}{cc}
                1 & \sigma_3 \\ -\sigma_3 & -1
                \end{array}\right)~,~~~~
                                 \gamma^-=\left(\begin{array}{cc}
                                       1 & -\sigma_3 \\ \sigma_3 & -1
                                           \end{array}\right)\;.
\end{equation}
Useful identities for many calculations are $\gamma^+
\gamma^- \gamma^+=4 \gamma^+$, and $\gamma^- \gamma^+ \gamma^-=4
\gamma^-$.

The operator that projects onto the dynamical fermion degree of
freedom is
\begin{equation}
\Lambda_+= {1 \over 2} \gamma^0 \gamma^+={1 \over 4} \gamma^-
\gamma^+= {1 \over 2} \left(\begin{array}{cc} 1 & \sigma_3 \\ \sigma_3
& 1 \end{array}\right) \;,
\end{equation}
and the complement projection operator is
\begin{equation}
\Lambda_-= {1 \over 2} \gamma^0 \gamma^-={1 \over 4} \gamma^+
\gamma^-= {1 \over 2}
\left(\begin{array}{cc} 1 & -\sigma_3 \\ -\sigma_3 & 1
                                         \end{array}\right) \;.
\end{equation}

The Dirac spinors $u(p,\sigma)$ and $v(p,\sigma)$ satisfy
\begin{equation}
(\pslash-m) u(p,\sigma)=0 \;,\;\;\; (\pslash+m) v(p,\sigma)=0 \;,
\end{equation}
and,
\begin{equation}
\overline{u}(p,\sigma)u(p,\sigma')=-\overline{v}(p,\sigma)
v(p,\sigma')=2 m \delta_{\sigma \sigma'}
\;,
\end{equation}
\begin{equation}
\overline{u}(p,\sigma) \gamma^\mu u(p,\sigma')=
\overline{v}(p,\sigma) \gamma^\mu v(p,\sigma')=
2 p^\mu \delta_{\sigma \sigma'}
\;,
\end{equation}
\begin{equation}
\sum_{\sigma=\pm {1 \over 2}} u(p,\sigma) \overline{u}(p,\sigma)
= \pslash+m \;,\;\;\; \sum_{\sigma=\pm {1 \over 2}}
v(p,\sigma) \overline{v}(p,\sigma)= \pslash-m
\;.
\end{equation}

There are only two physical gluon (photon) polarization vectors,
$\epsilon_{1\perp}$ and $\epsilon_{2\perp}$; but it is sometimes
convenient (and dangerous once covariance and gauge invariance are
violated) to use $\epsilon^\mu$, where
\begin{equation}
\epsilon^+=0 \;,\;\;\;\epsilon^-={2 {\bf q}_\perp \cdot \epsilon_\perp
\over q^+} \;.
\end{equation}
It is often possible to avoid using an explicit
representation for $\epsilon_\perp$, but completeness relations are
required,
\begin{equation}
\sum_{\lambda} \epsilon^\mu_\perp(\lambda)
\epsilon^{*\nu}_\perp(\lambda) = -g_\perp^{\mu \nu} \;,
\end{equation}
so that,
\begin{equation}
\sum_{\lambda} \epsilon^\mu(\lambda) \epsilon^{*\nu}(\lambda) =
-g_\perp^{\mu \nu} + {1 \over q^+} \bigl(\eta^\mu q_\perp^\nu +
\eta^\nu q_\perp^\mu \bigr) + { {\bf q}_\perp^2 \over (q^+)^2}
\eta^\mu \eta^\nu \;,
\end{equation}
where $\eta^+=\eta^1=\eta^2=0$ and $\eta^-=2$.  One often
encounters diagrammatic rules in which the gauge propagator is written
so that it looks covariant; but this is dangerous in loop calculations
because such expressions require one to add and subtract terms that
contain severe infrared divergences.

The QCD Lagrangian density is
\begin{equation}
{\cal L}=-{1 \over 2} Tr F^{\mu \nu} F_{\mu \nu} + \overline{\psi}
\Bigl(i \Dslash -m\Bigr)\psi \;,
\end{equation}
where $F^{\mu \nu}=\partial^\mu A^\nu-\partial^\nu A^\mu+i g
\bigl[A^\mu,A^\nu\bigr]$ and $i D^\mu=i \partial^\mu-g A^\mu$.  The
SU(3) gauge fields are $A^\mu=\sum_a A^\mu_a T^a$, where $T^a$ are
one-half the Gell-Mann matrices, $\lambda^a$,  and satisfy $Tr~ T^a
T^b= 1/2~ \delta^{ab}$ and $\bigl[T^a,T^b\bigr]=i f^{abc} T^c$.

The dynamical fermion degree of freedom is $\psi_+=\Lambda_+ \psi$,
and this can be expanded in terms of plane wave creation and
annihilation operators at $x^+=0$,
\begin{equation}
\psi_+^r(x)=\sum_{\sigma=\pm 1/2} \int_{k^+>0}
{dk^+ d^2k_\perp \over 16\pi^3 k^+}
\Bigl[b^r(k,\sigma) u_+(k,\sigma) e^{-ik\cdot x} +
d^{r\dagger}(k,\sigma) v_+(k,\sigma)
e^{ik\cdot x}\Bigr] \;,
\end{equation}
where these field operators satisfy
\begin{equation}
\Bigl\{\psi_+^r(x),\psi_+^{s\dagger}(y)\Bigr\}_{x^+=y^+=0}=\Lambda_+
\delta_{rs} \delta^3(x-y) \;,
\end{equation}
and the creation and annihilation operators satisfy
\begin{equation}
\Bigl\{b^r(k,\sigma),b^{s\dagger}(p,\sigma')\Bigr\}=
\Bigl\{d^r(k,\sigma),d^{s\dagger}(p,\sigma')\Bigr\}
= 16\pi^3 k^+ \delta_{rs} \delta_{\sigma \sigma'} \delta^3(k-p) \;.
\end{equation}
The indices $r$ and $s$ refer to SU(3) color.  In general,
when momenta are listed without specification of components, as in
$\delta^3(p)$, I am referring to $p^+$ and ${\bf p}_\perp$.

The transverse dynamical gluon field components can also be expanded
in terms of plane wave creation and annihilation operators,
\begin{equation}
A_\perp^{ic}(x)=\sum_\lambda\int_{k^+>0}
{dk^+ d^2k_\perp \over 16\pi^3 k^+}
\Bigl[a^c(k,\lambda) \epsilon_\perp^i(\lambda) e^{-ik\cdot x}
+a^{c\dagger}(k,\lambda) \epsilon_\perp^{i*}(\lambda) e^{ik\cdot x}
\Bigr] \;.
\end{equation}
The superscript $i$ refers to the transverse dimensions $x$
and $y$, and the superscript $c$ is for SU(3) color.  If required the
physical polarization vector can be represented
\begin{equation}
{\bf \epsilon}_\perp(\uparrow)=-{1 \over \sqrt{2}} (1,i) \;,\;\;\;
{\bf \epsilon}_\perp(\downarrow)={1 \over \sqrt{2}} (1,-i) \;.
\end{equation}
The quantization conditions are
\begin{equation}
\Bigl[A^{ic}_{\perp}(x),\partial^+ A^{jd}_{\perp}(y)
\Bigr]_{x^+=y^+=0} =i \delta^{ij} \delta_{cd} \delta^3(x-y) \;,
\end{equation}
\begin{equation}
\Bigl[a^c(k,\lambda),a^{d\dagger}(p,\lambda')\Bigr] = 16\pi^3 k^+
\delta_{\lambda \lambda'} \delta_{cd} \delta^3(k-p) \;.
\end{equation}

The classical equations for $\psi_-=\Lambda_- \psi$ and $A^-$ do
not involve time-derivatives, so these variables can be eliminated in
favor of dynamical degrees of freedom.  This formally yields
\begin{eqnarray}
\psi_-&=&{1 \over i\partial^+} \Bigl[i {\bf \alpha}_\perp \cdot {\bf
D}_\perp+\beta m\Bigr] \psi_+ \nonumber \\
&=&\psit_- - {g \over i\partial^+} {\bf \alpha}_\perp \cdot {\bf
A}_\perp \psi_+ \;,
\end{eqnarray}
where the variable $\psit_-$ is defined on the second
line to separate the interaction-depen\-dent part of $\psi_-$; and
\begin{eqnarray}
A^{a-}&=&{2 \over i\partial^+} i {\bf \partial}_\perp
\cdot {\bf A}^a_\perp +
{2 i g f^{abc} \over (i\partial^+)^2}\Biggl\{\bigl(i\partial^+
A^{bi}_\perp\bigr)A^{ci}_\perp +2 \psi_+^\dagger T^a \psi_+\Biggr\}
\nonumber \\
&=&\At^{a-}+{2 i g f^{abc} \over (i\partial^+)^2}
\Biggl\{\bigl(i\partial^+ A^{bi}_\perp\bigr) A^{ci}_\perp
+2 \psi_+^\dagger T^a \psi_+ \Biggr\}
\;,
\end{eqnarray}
where the variable $\At^-$ is defined on the second line to
separate the interaction-depen\-dent part of $A^-$.

Given these replacements, we can follow a canonical procedure to
determine the Hamiltonian.  This path is full of difficulties that
I ignore, because ultimately I will use coupling coherence to refine
the definition of the Hamiltonian and determine the non-canonical
interactions that are inevitably produced by the violation of explicit
covariance and gauge invariance.  For my purposes it is sufficient to
write down a Hamiltonian that can serve as a starting point:
\begin{equation}
H=H_0+V \;,
\end{equation}
\begin{eqnarray}
H_0&=&\int d^3x \Bigl\{ Tr\bigl(\partial^i_\perp A^j_\perp
\partial^i_\perp A^j_\perp\bigr)+\psi_+^\dagger \bigl(i \alpha^i_\perp
\partial^i_\perp+\beta m\bigr) \psi_+\Bigr\} \nonumber \\
&=&\sum_{colors} \int {dk^+ d^2k_\perp \over 16\pi^3 k^+} \Biggl\{
\sum_\lambda {{\bf k}_\perp^2 \over k^+} a^\dagger(k,\lambda)
a(k,\lambda) \nonumber \\
&&~~~~~~~~~~~+ \sum_\sigma {{\bf k}_\perp^2+m^2 \over k^+}
\Bigl(b^\dagger(k,\sigma) b(k,\sigma)+d^\dagger(k,\sigma) d(k,\sigma)
\Bigr) \Biggr\} \;.
\end{eqnarray}
In the last line the `self-induced inertias' ({\it i.e.},
one-body operators produced by normal-ordering $V$) are not included.
It is difficult to regulate the field contraction encountered when
normal-ordering in a manner exactly consistent with the cutoff
regulation of contractions encountered later.  Coupling coherence
avoids this issue and produces the correct one-body counterterms with
no discussion of normal-ordering required.

The interactions are complicated and are most easily
written using the variables, $\psit=\psit_- + \psi_+$,
and $\At$, where $\At^+=0$, $\At^-$ is defined above, and
$\At^i_\perp=A^i_\perp$.  Using these variables we have
\begin{eqnarray}
V&=\int d^3x\Biggl\{& g \overline{\psit} \gamma_\mu \At^\mu \psit +
2 g~ Tr\Bigl(i\partial^\mu \At^\nu \Bigl[
\At_\mu, \At_\nu \Bigr] \Bigr) \nonumber \\
&&-{g^2 \over 2} Tr\Bigl( \Bigl[\At^\mu,\At^\nu\Bigr] \Bigl[
\At_\mu, \At_\nu \Bigr] \Bigr) + g^2 \overline{\psit} \gamma_\mu
\At^\mu {\gamma^+ \over 2i\partial^+} \gamma_\nu \At^\nu \psit
\nonumber \\
&&+{g^2 \over 2} \overline{\psi} \gamma^+ T^a \psi {1 \over
(i\partial^+)^2} \overline{\psi} \gamma^+ T^a \psi \nonumber \\
&&-g^2 \overline{\psi} \gamma^+ \Biggl({1 \over (i\partial^+)^2}
\Bigl[i\partial^+ \At^\mu, \At_\mu \Bigr] \Biggr) \psi \nonumber \\
&&+g^2 ~Tr \Biggl(\Bigl[i\partial^+ \At^\mu, \At_\mu \Bigr] {1 \over
(i\partial^+)^2} \Bigl[i\partial^+ \At^\nu, \At_\nu \Bigr] \Biggr)
~~~\Biggr\}\;.
\end{eqnarray}
The commutators in this expression are SU(3) commutators
only.  The potential algebraic complexity of calculations becomes
apparent when one systematically expands every term in $V$ and
replaces:
\begin{equation}
\psit^- \rightarrow {1 \over i\partial^+}\Bigl(i {\bf \alpha}_\perp
\cdot {\bf \partial}_\perp+\beta m\Bigr) \psi_+ \;,
\end{equation}
\begin{equation}
\At^- \rightarrow {2 \over i\partial^+} i{\bf \partial}_\perp \cdot
{\bf A}_\perp \;;
\end{equation}
and then expands $\psi_+$ and ${\bf A}_\perp$ in terms of
creation and annihilation operators.  It rapidly becomes evident that
one should avoid such explicit expansions if possible.

\section{LIGHT-FRONT QED}

In this section I will follow the strategy outlined in the first
section to compute the positronium spectrum.  I will detail the
calculation through the leading order Bohr results\site{brazil} and
indicate how higher order calculations proceed.\site{BJ97a,BJ97c} 

The first step is to compute a renormalized cutoff Hamiltonian as a
power series in the coupling $e$. Starting with the canonical
Hamiltonian as a `seed,' this is done with the similarity
renormalization group\site{glazek1,glazek2} and coupling
coherence.\site{coupcoh,perryrg}  The result is an apparently unique
perturbative series,
\begin{equation}
H^\Lambda_N=h_0 + e_\Lambda h_1 +e_\Lambda^2 h_2 +
\cdot\cdot\cdot + e_\Lambda^N h_N \;.
\end{equation}

Here $e_\Lambda$ is the running coupling constant, and all remaining
dependence on $\Lambda$ in the operators $h_i$ must be explicit. In
principle I must also treat $m_\Lambda$, the electron running mass, as
an independent function of $\Lambda$; but this will not affect the
results to the order I compute here. We must calculate the
Hamiltonian to a fixed order, and systematically improve the
calculation later by including higher order terms.

Having obtained the Hamiltonian to some order in $e$, the next step is
to split it into two parts,
\begin{equation}
H^\Lambda=\H_0+\V \;.
\end{equation}
As discussed before, $\h0$ must be accurately solved
non-perturbatively, producing a zeroth order approximation for the
eigenvalues and eigenstates.  The greatest ambiguities in the
calculation appear in the choice of $\h0$, which requires one of
science's most powerful computational tools, trial and error.

In QED and QCD I {\it assume} that all interactions in $\H_0$ preserve
particle number, with all interactions that involve particle creation
and annihilation in $\V$.  This assumption is consistent with the
original hypothesis that a constituent picture will emerge, but it
should emerge as a valid approximation.

The final step before the loop is repeated, starting with a more
accurate approximation for $H^\Lambda$, is to compute corrections from
$\V$ in bound state perturbation theory.  There is no reason
to compute these corrections to arbitrarily high order, because the
initial Hamiltonian contains errors that limit the accuracy we can
obtain in bound state perturbation theory.

In this section I: (i) compute $H^\Lambda$ to ${\cal O}(e^2)$, (ii)
assume the cutoff is in the range $\alpha m^2 < \Lambda^2 < \alpha^2
m^2$ for non-perturbative analyses, (iii) include the most infrared
singular two-body interactions in $\H_0$, and (iv) estimate the
binding energy for positronium to ${\cal O}(\alpha^2 m)$.

Since $\H_0$ is assumed to include interactions that preserve particle
number, the zeroth order positronium ground state will be a pure
electron-positron state.  We only need  one- and two-body
interactions; {\it i.e.}, the electron self-energy and the
electron-positron interaction.  The canonical interactions can be
found in Eq. (113), and the second-order change in the Hamiltonian is
given in Eq. (57). The shift due to the bare electron mixing with
electron-photon states to lowest order (see Figure 1) is
\begin{equation}
\delta \Sigma_p=e^2 \int {dk_1^+ d^2k_{1\perp} \over 16 \pi^3}
{\theta(k_1^+) \theta(k_2^+) \over k_1^+ k_2^+} {\overline{u} (p)
D_{\mu\nu}(k_1) \gamma^\mu \bigl(\kslash_2+m\bigr) \gamma^\nu u(p)
\over p^--k_1^--k_2^-} \;,
\end{equation}
where,
\begin{equation} k_2^+=p^+-k_1^+ \;,\;\;\;{\bf k}_{2\perp}={\bf
p}_\perp-{\bf k}_{1\perp} \;,
\end{equation}
\begin{equation}
k_i^-={k_{i\perp}^2 + m_i^2 \over k_i^+} \;,
\end{equation}
\begin{equation}
D_{\mu\nu}(k)=-g_{\perp \mu \nu} + {k_\perp^2 \over (k^+)^2} \eta_\mu
\eta_\nu + {1 \over k^+} \Bigl(k_{\perp \mu} \eta_\nu + k_{\perp \nu}
\eta_\mu\Bigr) \;,
\end{equation}
\begin{equation}
\eta_\mu a^\mu = a^+ \;.
\end{equation}
I have not yet displayed the cutoffs.  To evaluate the
integrals it is easiest to use Jacobi variables $x$ and ${\bf s}$ for
the relative electron-photon motion,
\begin{equation}k_1^+=x p^+\;,\;\;\;{\bf k}_{1\perp}=x {\bf
p}_\perp+{\bf s} \;,
\end{equation}
which implies
\begin{equation}k_2^+=(1-x) p^+\;,\;\;\;{\bf k}_{2\perp}=(1-x) {\bf
p}_\perp-{\bf s} \;.
\end{equation}

The second-order change in the electron self-energy becomes
\begin{eqnarray}
\delta \Sigma_p &=& -{e_\Lambda^2 \over p^+} \int {dx d^2s \over
16\pi^3}
\theta\Biggl(y \Lambda_0^2-{s^2 +x^2 m^2 \over x (1-x)}\Biggr)
\theta\Biggl({s^2 +x^2 m^2 \over x (1-x)}-y \Lambda_1^2\Biggr)
\nonumber \\
&&~~~\Biggl({1 \over s^2+x^2 m^2}\Biggr)~
\overline{u}(p,\sigma) \Biggl\{ 2 (1-x) \pslash -2 m + \nonumber \\
&&~~~~~~~~~~~~~~~~~~~~~~{\gamma^+ \over p^+}
\Biggl[ {2 s^2 \over x^2}+{2s^2 \over 1-x}+{x (2-x) m^2 \over 1-x}
\Biggr]
\Biggr\} u(p,\sigma) \;,
\end{eqnarray}
where $y=p^+/P^+$.

It is straightforward in this case to determine the self-energy
required by coupling coherence.  Since the electron-photon coupling
does not run until third order, to second order the self-energy must
exactly reproduce itself with $\Lambda_0 \rightarrow \Lambda_1$.
For the self-energy to be finite we must assume that $\delta \Sigma$
reduces a positive self-energy, so that
\begin{eqnarray}
\Sigma^{\Lambda}_{coh}(p)&=& {e_\Lambda^2 \over p^+}
\int {dx d^2s \over 16\pi^3} \theta\bigl(yx-\epsilon\bigr)
\theta\Biggl(y \Lambda^2-{s^2 +x^2 m^2 \over x (1-x)}\Biggr)
\Biggl({1 \over s^2+x^2 m^2}\Biggr) \nonumber
\\&&
\overline{u}(p,\sigma)
\Biggl\{ 2 (1-x) \pslash -2 m +{\gamma^+ \over p^+}
\Biggl[ {2 s^2 \over x^2}+{2s^2 \over 1-x}+
{x (2-x) m^2 \over 1-x} \Biggr]
\Biggr\} u(p,\sigma) \nonumber \\
&=&{e_\Lambda^2 \over 8\pi^2 p^+}
\Biggl\{2 y \Lambda^2 \ln\Biggl(
{ y^2 \Lambda^2 \over (y\Lambda^2 + m^2)\epsilon}\Biggr)
-{3 \over 2} y \Lambda^2+{1 \over 2}
{ym^2 \Lambda^2 \over y \Lambda^2+m^2} \nonumber
\\ &&~~~~~~~~~~~~~~~~~~~~~~~~+ 3 m^2
\ln\Biggl( {m^2 \over y \Lambda^2 + m^2} \Biggr) \Biggl\} + {\cal
O}(\epsilon/y) \;.
\end{eqnarray}

I have been forced to introduce {\it a second cutoff},
\begin{equation}
x p^+ > \epsilon P^+ \;,
\end{equation}
because after the ${\bf s}$ integration is completed we are
left with a logarithmically divergent $x$ integration.  Other choices
for this second infrared cutoff are possible and lead to similar
results.  This second cutoff must be taken to zero and no new
counterterms can be added to the Hamiltonian, so all divergences must
cancel before it is taken to zero.

The electron and photon (quark and gluon) `mass' operators, are a
function of a longitudinal momentum scale introduced by the cutoff,
and there is an exact scale invariance required by longitudinal boost
invariance.  Here I mean by `mass operator' the one-body operator
when the transverse momentum is zero, even though this does not agree
with the free mass operator because it includes longitudinal momentum
dependence.  The cutoff violates boost invariance and the mass
operator is required to restore this symmetry.

We must interpret this new infrared divergence, because we have no
choice about whether it is in the Hamiltonian if we use coupling
coherence.  We can only choose between putting the divergent operator
in $\H_0$ or in $\V$.  I make different choices in QED and QCD, and
the arguments are based on physics.

The {\it divergent} electron `mass' is a complete lie.  We encounter
a term proportional to $e_\Lambda^2 \Lambda^2 \ln(1/\epsilon)/P^+$
when the scale is $\Lambda$; however, we can reduce this scale as far
as we please in perturbation theory.  Photons are massless, so the
electron will continue to dress itself with small-x photons to
arbitrarily small $\Lambda$. Since I believe that this divergent
self-energy is exactly canceled by mixing with small-x photons, and
that this mixing can be treated perturbatively in QED, I simply put
the divergent electron self-energy in $\V$, which is treated
perturbatively.

There are two time-ordered diagrams involving photon exchange
between an electron with initial momentum $p_1$ and final momentum
$p_2$, and a positron with initial momentum $k_1$ and final momentum
$k_2$.  These are shown in Figure 4, along with the instantaneous
exchange diagram. Using Eq. (57), we find the required matrix element
of $\delta H$,
\begin{eqnarray}
\delta H &=& -{e^2 \over q^+} D_{\mu
\nu}(q) \overline{u}(p_2,\sigma_2) \gamma^\mu u(p_1,\sigma_1)
\overline{v}(k_1,\lambda_1) \gamma^\nu v(k_2,\lambda_2) \nonumber \\
&&~~\theta\bigl(|q^+|-\epsilon P^+\bigr)
\theta\Biggl({\Lambda_0^2 \over P^+} - \mid p_1^- -p_2^-
-q^- \mid\Biggr) \theta\Biggl({\Lambda_0^2 \over P^+} -
\mid k_2^- -k_1^- -q^- \mid\Biggr) \nonumber \\
&&~~\Biggl[ {\theta\bigl(|p_1^- -p_2^- -q^-| -\Lambda_1^2 / {\cal P}^+
\bigr)
\;\; \theta\bigl(|p_1^- -p_2^- -q^-|- |k_2^- -k_1^- -q^-| \bigr)
\over p_1^- -p_2^- -q^-} \nonumber \\ &&~~~~
+{\theta\bigl(|k_2^- -k_1^- -q^-| - \Lambda_1^2 / {\cal P}^+ \bigr)
\;\; \theta\bigl( |k_2^- -k_1^- -q^-| - |p_1^- -p_2^- -q^-| \bigr)
\over k_2^- -k_1^- -q^-} \Biggr]
\nonumber \\
&&~~~~~~~~~~~\theta\Biggl(\Lone-\mid p_1^-+k_1^--p_2^--k_2^- \mid
\Biggr) \;, 
\end{eqnarray}
where $q^+=p_1^+-p_2^+\;,\;\;{\bf q}_\perp={\bf p}_{1\perp}
- {\bf p}_{2\perp}$, and $q^-=q_\perp^2/q^+$.

I have used the second cutoff on longitudinal momentum that I was
forced to introduce when computing the change in the self-energy.  We
will see in the section on confinement that it is essential to include
this cutoff everywhere consistently.  In QED this point is not
immediately important, because all infrared singular interactions,
including the infrared divergent self-energy, are put in $\V$ and
treated perturbatively.  Divergences from higher orders in $\V$
cancel.\site{BJ97a,BJ97c}

\begin{figure}
\vskip6mm
\begin{center}
\epsfig{file=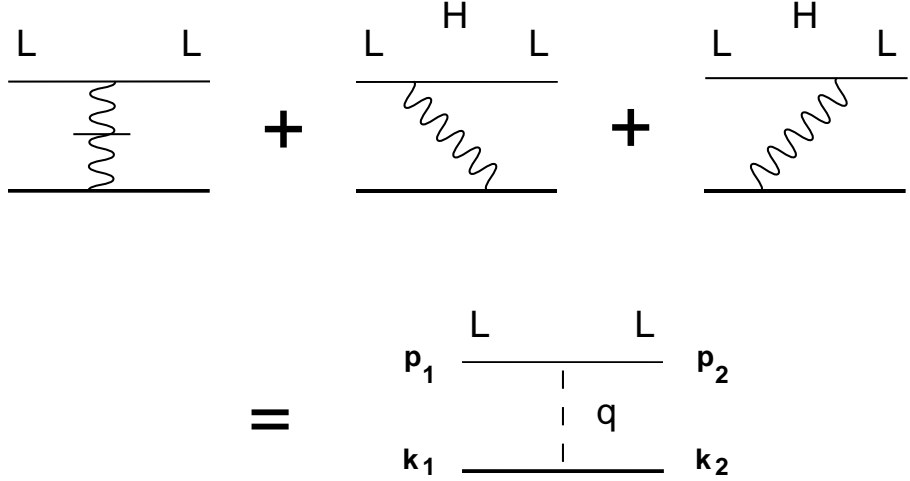,width=12cm}
\end{center} 
\vskip-3mm
\caption{Effective two-body interaction between low energy
constituents resulting from: (i) the canonical instantaneous exchange
interaction, and (ii) the elimination of direct coupling between
low energy two-body states and high energy states containing an
additional gauge particle.}
\label{fig:fig4}
\end{figure}

To determine the interaction that must be added to the Hamiltonian to
maintain coupling coherence, we must again find an interaction that
when added to $\delta V$ reproduces itself with $\Lambda_0
\rightarrow \Lambda_1$ everywhere.  The coupling coherent interaction
generated by the first terms in $\delta H$ are not uniquely
determined at this order. There is some ambiguity because we can
obtain coupling coherence either by having $\delta H$ increase the
strength of an operator by adding additional phase space strength, or
we can have $\delta H$ reduce the strength of an operator by
subtracting phase space strength.  The ambiguity is resolved in
higher orders, so I will simply state the result.  If an
instantaneous photon-exchange interaction is present in $H$, 
$\delta H$ cancels part of this marginal operator and increases
the strength of a new photon-exchange interaction. This new
interaction reproduces the effects of high energy photon exchange
removed by the cutoff.  The result is
\begin{eqnarray}
V_{coh}^{\Lambda} &=& - {e_{\Lambda}^2 \over q^+}  D_{\mu
\nu}(q) \overline{u}(p_2,\sigma_2) \gamma^\mu u(p_1,\sigma_1)
\overline{v}(k_1,\lambda_1) \gamma^\nu v(k_2,\lambda_2) \nonumber \\
&&~~~~~~~\theta\bigl(|q^+|-\epsilon P^+\bigr)~
\theta\Biggl(\Lam-\mid p_1^-+k_1^--p_2^--k_2^- \mid
\Biggr) \nonumber \\
&&~~\Biggl[ {\theta\bigl(|p_1^- -p_2^- -q^-| -\Lambda^2 / {\cal P}^+
\bigr)
\;\; \theta\bigl(|p_1^- -p_2^- -q^-|- |k_2^- -k_1^- -q^-| \bigr)
\over p_1^- -p_2^- -q^-} \nonumber \\ &&~~~~
+{\theta\bigl(|k_2^- -k_1^- -q^-| - \Lambda^2 / {\cal P}^+ \bigr)
\;\; \theta\bigl( |k_2^- -k_1^- -q^-| - |p_1^- -p_2^- -q^-| \bigr)
\over k_2^- -k_1^- -q^-} \Biggr]  \;. \nonumber \\
\end{eqnarray}

This matrix element exactly reproduces photon exchange above the
cutoff.  The cutoff removes the direct coupling of electron-positron
states to electron-positron-pho\-ton states whose energy differs by more
than the cutoff, and coupling coherence dictates that the result of
this mixing should be replaced by a direct interaction between the
electron and positron.  We could obtain this result by much simpler
means at this order by simply demanding that the Hamiltonian produce
the `correct' scattering amplitude at $\order(e^2)$ with the cutoffs
in place.  Of course, this procedure requires us to provide the
`correct' amplitude, but this is easily done in perturbation theory.

$V^\Lambda_{coh}$ is non-canonical, and we will see that it is
responsible for producing the Coulomb interaction. We need some
guidance to decide which irrelevant operators are most important. We
find {\it a posteriori} that differences of external transverse
momenta, and differences of external longitudinal momenta are both
proportional to $\alpha$.  This allows us to identify the dominant
operators by expanding in powers of these {\it implicit} powers of
$\alpha$.  This indicates that it is the {\it most infrared singular}
part of $V_{coh}$ that is important.  As explained above, this
operator receives substantial strength only from the exchange of
photons with small longitudinal momentum; so we expect inverse $q^+$
dependence to indicate `strong' interactions between low energy
pairs.  So the part of $V_{coh}$ that is included in $\H_0$ is
\begin{eqnarray}
\tilde{V}_{coh}^{\Lambda} &=& - e_{\Lambda}^2
{q_\perp^2 \over (q^+)^3}~
\overline{u}(p_2,\sigma_2) \gamma^+ u(p_1,\sigma_1)
\overline{v}(k_1,\lambda_1) \gamma^+ v(k_2,\lambda_2) \nonumber \\
&&~~~~~~~\theta\bigl(|q^+|-\epsilon P^+\bigr)~
\theta\Biggl(\Lam-\mid p_1^-+k_1^--p_2^--k_2^- \mid
\Biggr) \nonumber \\
&&~~\Biggl[ {\theta\bigl(|p_1^- -p_2^- -q^-| -\Lambda^2 / {\cal P}^+ \bigr)
\;\; \theta\bigl(|p_1^- -p_2^- -q^-|- |k_2^- -k_1^- -q^-| \bigr)
\over p_1^- -p_2^- -q^-} \nonumber \\ &&~~~~
+{\theta\bigl(|k_2^- -k_1^- -q^-| - \Lambda^2 / {\cal P}^+ \bigr)
\;\; \theta\bigl( |k_2^- -k_1^- -q^-| - |p_1^- -p_2^- -q^-| \bigr)
\over k_2^- -k_1^- -q^-} \Biggr]   \nonumber \\
&=& - 4 e_{\Lambda}^2 \sqrt{p_1^+ p_2^+ k_1^+ k_2^+}
{q_\perp^2 \over (q^+)^3} \delta_{\sigma_1 \sigma_2}
\delta_{\lambda_1 \lambda_2} \nonumber \\
&&~~~~~~~\theta\bigl(|q^+|-\epsilon P^+\bigr)~
\theta\Biggl(\Lam-\mid p_1^-+k_1^--p_2^--k_2^- \mid
\Biggr) \nonumber \\
&&~~\Biggl[ {\theta\bigl(|p_1^- -p_2^- -q^-| -\Lambda^2 / {\cal P}^+ \bigr)
\;\; \theta\bigl(|p_1^- -p_2^- -q^-|- |k_2^- -k_1^- -q^-| \bigr)
\over p_1^- -p_2^- -q^-} \nonumber \\ &&~~~~
+{\theta\bigl(|k_2^- -k_1^- -q^-| - \Lambda^2 / {\cal P}^+ \bigr)
\;\; \theta\bigl( |k_2^- -k_1^- -q^-| - |p_1^- -p_2^- -q^-| \bigr)
\over k_2^- -k_1^- -q^-} \Biggr]  \;. \nonumber \\
\end{eqnarray}

The Hamiltonian is almost complete to second order in the
electron-positron sector, and only the instantaneous photon exchange
interaction must be added.  The matrix element of this interaction is
\begin{eqnarray}
V_{instant} &=& - e_{\Lambda}^2 \Biggl({1 \over q^+}\Biggr)^2
\overline{u}(p_2,\sigma_2)
\gamma^+ u(p_1,\sigma_1) \overline{v}(k_1,\lambda_1) \gamma^+
v(k_2,\lambda_2) \nonumber \\
&&~~~~~~~~~\times
\theta\Biggl(\Lam-\mid p_1^-+k_1^--p_2^--k_2^- \mid
\Biggr) \nonumber \\
&=& - 4 e_{\Lambda}^2 \sqrt{p_1^+ p_2^+ k_1^+ k_2^+}  \Biggl({1 \over
q^+}\Biggr)^2 \delta_{\sigma_1 \sigma_2}
\delta_{\lambda_1 \lambda_2} \nonumber \\
&&~~~~~~~~~~~~\times
\theta\Biggl(\Lam-\mid p_1^-+k_1^--p_2^--k_2^- \mid
\Biggr)
\;.
\end{eqnarray}
The only cutoff that appears is the cutoff directly run by
the similarity transformation that prevents the initial and final
states from differing in energy by more than $\Lambda^2/P^+$.

This brings us to a final subtle point.  Since there are no cutoffs in
$V_{instant}$ that directly limit the momentum exchange, the matrix
element diverges as $q^+ \rightarrow 0$.  Consider $\tilde{V}_{coh}$
in this same limit,
\begin{eqnarray}
\tilde{V}_{coh}^{\Lambda} &\rightarrow&
4 e_{\Lambda}^2 \sqrt{p_1^+ p_2^+ k_1^+ k_2^+} \Biggl({1 \over
q^+}\Biggr)^2  \delta_{\sigma_1 \sigma_2}
\delta_{\lambda_1 \lambda_2} \nonumber \\
&&~~~~ \times \theta\Biggl(\mid q^- \mid - {\Lambda^2 \over P^+}
\Biggr)  ~ \theta\Biggl(\Lam-\mid p_1^-+k_1^--p_2^--k_2^- \mid \Biggr)
\;.
\end{eqnarray}
This means that as $q^+ \rightarrow 0$, $V_{coh}$ partially
screens $V_{instant}$, leaving the original operator multiplied by
$\theta(\Lambda^2/P^+-|q^-|)$.  However, even after this partial
screening, the matrix elements of the remaining part of $V_{instant}$
between bound states diverge and we must introduce the same infrared
cutoff used for the self-energy to regulate these divergences. This is
explicitly shown in the section on confinement. However, all
divergences from $V_{instant}$ are exactly canceled by the exchange
of massless photons, which persists to arbitrarily small cutoff.  This
cancellation is exactly analogous to the cancellation of the infrared
divergence of the self-energy, and will be treated in the same way.
The portion of $V_{instant}$ that is not canceled by
$\tilde{V}_{coh}$ will be included in $\V$, the perturbative part of
the Hamiltonian.  We will not encounter this interaction until we also
include photon exchange below the cutoff perturbatively, so all
infrared divergences should cancel in this bound state perturbation
theory.  I repeat that this is not guaranteed for arbitrary choices of
$\H_0$, and we are not free to simply cancel these divergent
interactions with counterterms because coupling coherence completely
determines the Hamiltonian.

We now have the complete interaction that I include in $\H_0$. 
Letting $\H_0=h_0+\v_0$, where $h_0$ is the free hamiltonian, I add
parts of $V_{instant}$ and $V_{coh}$ to obtain
\begin{eqnarray} 
\v_0 &=& - 4 e_{\Lambda}^2 \sqrt{p_1^+ p_2^+ k_1^+ k_2^+}
\delta_{\sigma_1 \sigma_2} \delta_{\lambda_1 \lambda_2}
\theta\Biggl(\Lam-\mid p_1^-+k_1^--p_2^--k_2^- \mid
\Biggr) \times \nonumber \\ 
&& \Biggl\{ 
\Biggl[ {q_\perp^2 \over (q^+)^3} { 1 \over p_1^- -p_2^- -q^-}
+ {1 \over (q^+)^2} \Biggr]  \theta\bigl(|p_1^- -p_2^- -q^-|
-\Lambda^2 / {\cal P}^+ \bigr) \nonumber \\
&&~~~~~~~~~~~~~~~~~~~~~~~~~~~~ \times \theta\bigl(|p_1^- -p_2^-
-q^-|-  |k_2^- -k_1^- -q^-| \bigr) \nonumber \\ 
&&+
\Biggl[ {q_\perp^2 \over (q^+)^3} {1 \over k_2^- -k_1^- -q^-} 
+ {1 \over (q^+)^2} \Biggr]   \theta\bigl(|k_2^- -k_1^- -q^-| 
- \Lambda^2 / {\cal P}^+ \bigr) \nonumber \\
&&~~~~~~~~~~~~~~~~~~~~~~~~~~~~ \times \theta\bigl( |k_2^- -k_1^-
-q^-| - |p_1^- -p_2^- -q^-| \bigr) \Biggr\} \;.
\end{eqnarray}

In order to present an analytic analysis I will make assumptions that
can be justified $a~ posteriori$.  First I will assume that the
electron and positron momenta can be arbitrarily large, but that in
low-lying states their relative momenta satisfy
\begin{equation}
|{\bf p}_\perp-{\bf k}_\perp| \ll m \;,
\end{equation}
\begin{equation}
|p^+-k^+| \ll p^++k^+ \;.
\end{equation}
It is essential that the condition for longitudinal momenta
not involve the electron mass, because masses have the scaling
dimensions of transverse momenta and not longitudinal momenta.  As
above, I use $p$ for the electron momenta and $k$ for the positron
momenta.  To be even more specific, I will assume that
\begin{equation}
|{\bf p}_\perp-{\bf k}_\perp| \sim \alpha m \;,
\end{equation}
\begin{equation}
|p^+-k^+| \sim \alpha (p^++k^+) \;.
\end{equation}
This allows us to use power counting to evaluate the
perturbative strength of operators for small coupling, which may prove
essential in the analysis of QCD.\site{nonpert} Note that these
conditions allow us to infer
\begin{equation}
|{\bf p}_{1\perp}-{\bf p}_{2\perp}| \sim \alpha m \;,
\end{equation}
\begin{equation}
|p_1^+-p_2^+| \sim \alpha (p^++k^+) \;.
\end{equation}

Given these order of magnitude estimates for momenta, we can
drastically simplify the free energies in the kinetic energy operator
and the energy denominators in $\v_0$.  We can use transverse boost
invariance to choose a frame in which
\begin{equation}
p_i^+=y_i P^+ \;,\;\;p_{i\perp}=\kappa_i \;,\;\;\;\;k_i^+=(1-y_i)P^+
\;,\;\;k_{i\perp}=-\kappa_i \;,
\end{equation}
so that
\begin{eqnarray}
P^+(p_1^--p_2^--q^-) &=& {\kappa_1^2+m^2 \over y_1} - {\kappa_2^2+m^2
\over y_2} - {(\kappa_1-\kappa_2)^2 \over y_1-y_2} \nonumber
\\&=&-4 m^2 (y_1-y_2) - {(\kappa_1-\kappa_2)^2 \over y_1-y_2} +
\order(\alpha^2 m^2) \;.
\end{eqnarray}
To leading order all energy denominators are the same.
Each  energy denominator is $\order(\alpha m^2)$, which is large in
comparison to the binding energy we will find.  This is important,
because the bulk of the photon exchange that is important for the low
energy bound state involves intermediate states that have larger
energy than the differences in constituent energies in the region of
phase space where the wave function receives most of its strength.
This allows us to use a perturbative renormalization group to compute
the dominant effective interactions.

There are similar simplifications for all energy denominators.  After
making these approximations we find that the matrix element of
$\v_0$ is
\begin{eqnarray}
\v_0&=&4 e_\Lambda^2
\sqrt{y_1 y_2 (1-y_1) (1-y_2)} ~\theta\Bigl( 4 m^2
|y_1-y_2| + {(\kappa_1-\kappa_2)^2 \over |y_1-y_2|}
-\Lambda^2\Bigr) \nonumber \\
&&~~~~~\theta\Bigl(\Lambda^2-4\mid \kappa_1^2+4 m^2(1-2y_1)^2-
\kappa_2^2-4m^2(1-2y_2)^2 \mid\Bigr) \nonumber \\
&&~~~~~~~~~~~~~~~~~~~ {(\kappa_1-\kappa_2)^2
\over (y_1-y_2)^2} \Biggl[{1 \over 4 m^2 (y_1-y_2)^2+(\kappa_1 -
\kappa_2)^2}-{1 \over (\kappa_1-\kappa_2)^2}\Biggr] \nonumber \\
&=&-16 e_\Lambda^2 m^2 \sqrt{y_1 y_2 (1-y_1) (1-y_2)}~ \theta\Bigl( 4
m^2 |y_1-y_2| + {(\kappa_1-\kappa_2)^2
\over |y_1-y_2|}-\Lambda^2\Bigr) \nonumber
\\ &&~~~~\theta\Bigl(\Lambda^2-4\mid \kappa_1^2+4 m^2(1-2y_1)^2-
\kappa_2^2-4m^2(1-2y_2)^2 \mid\Bigr) \nonumber \\
&&~~~~~~~~~~~~~~~~~~~~~~\Biggl[
{1 \over 4 m^2 (y_1-y_2)^2+(\kappa_1-\kappa_2)^2} \Biggr] \;.
\end{eqnarray}
In principle the electron-positron annihilation graphs
should also be included at this order, but the resultant effective
interactions do not diverge as $q^+ \rightarrow 0$, so I include such
effects perturbatively in $\V$.

At this point we can complete the zeroth order analysis of positronium
using the state,
\begin{eqnarray}
|\Psi(P)\rangle &=& \sum_{\sigma \lambda} \int {dp^+ d^2p_\perp \over
16\pi^3 p^+} {dk^+ d^2k_\perp \over 16\pi^3 k^+} \sqrt{p^+ k^+}
16\pi^3 \delta^3(P-p-k) \nonumber \\
&&~~~~~~~~~~~~~~\phi(p,\sigma;k,\lambda) b^\dagger(p,\sigma)
d^\dagger(k,\lambda) |0\rangle \;,
\end{eqnarray}
where $\phi(p,\sigma;k,\lambda)$ is the wave function for
the relative motion of the electron and positron, with the
center-of-mass momentum being $P$.  We need to choose the longitudinal
momentum appearing in the cutoff, and I will use the natural scale
$P^+$.  The matrix element of $\H_0$ is
\begin{eqnarray}
&&\langle \Psi(P)|\H_0|\Psi(P')\rangle = 16\pi^3 \delta^3(P-P') \times
\nonumber \\
&&~~~\Biggl\{ \int {dy d^2\kappa \over 16\pi^3}
\Bigl[4m^2+4\kappa^2+4m^2(1-2y)^2\Bigr] |\phi(\kappa,y)|^2 \nonumber \\
&&~~~~~-16 e^2 m^2 \int {dy_1 d^2\kappa_1 \over 16\pi^3} {dy_2
d^2\kappa_2 \over 16\pi^3} \theta\Bigl(4 m^2|y_1-y_2| + {(\kappa_1 -
\kappa_2)^2 \over |y_1-y_2|}-\Lambda^2\Bigr) \nonumber \\
&&~~~~~~~~~~~\theta\Bigl(\Lambda^2-4\mid \kappa_1^2+4 m^2(1-2y_1)^2-
\kappa_2^2-4m^2(1-2y_2)^2 \mid\Bigr) \nonumber \\
&&~~~~~~~~~~~~~~~\Biggl[{1 \over 4m^2|y_1-y_2|^2+(\kappa_1-\kappa_2)^2}
\Biggr] \phi^*(\kappa_2,y_2) \phi(\kappa_1,y_1) \Biggr\} \;.
\end{eqnarray}
I have chosen a frame in which $P_\perp=0$ and used the
Jacobi coordinates defined above, and indicated only the electron
momentum in the wave function since momentum conservation fixes the
positron momentum.  I have also dropped the spin indices because the
interaction in $\H_0$ is independent of spin.

If we vary this expectation value subject to the constraint that the
wave function is normalized we obtain the equation of motion,
\begin{eqnarray}
M^2 \phi(\kappa_1,y_1) &=& (4 m^2 - 4 m E + E^2) \phi(\kappa_1,y_1)
\nonumber \\&=& \Bigl[4m^2+4\kappa_1^2+4m^2(1-2y_1)^2\Bigr]
\phi(\kappa_1,y_1) \nonumber \\
&&-16 e^2 m^2 \int {dy_2 d^2\kappa_2 \over 16\pi^3}
\theta\Bigl(4 m^2|y_1-y_2| + {(\kappa_1 -
\kappa_2)^2 \over |y_1-y_2|} -\Lambda^2\Bigr) \nonumber \\
&&~~~~~~\theta\Bigl(\Lambda^2-4\mid \kappa_1^2+4 m^2(1-2y_1)^2-
\kappa_2^2-4m^2(1-2y_2)^2 \mid\Bigr) \nonumber \\
&&~~~~~~~~~~\Biggl[{1 \over 4m^2|y_1-y_2|^2+(\kappa_1-\kappa_2)^2}
\Biggr] \phi(\kappa_2,y_2) \;.
\end{eqnarray}
$E$ is the binding energy, and we can drop the $E^2$ term
since it will be $\order(\alpha^4)$.

I do not think that it is possible to solve this equation analytically
with the cutoffs in place, and with the light-front kinematic
constraints $-1 \le 1-2y_i \le 1$.  In order to determine the binding
energy to leading order, we need to evaluate the regions of phase
space removed by the cutoffs.

If we want to find a cutoff for which the ground state is dominated by
the electron-positron component of the wave function, we need the first
cutoff to remove the important part of the electron-positron-photon
phase space.  Using the `guess' that $|\kappa|=\order(\alpha m)$ and
$1-2y=\order(\alpha)$, this requires
\begin{equation}
\Lambda^2<\alpha m^2 \;.
\end{equation}
On the other hand, we cannot allow the cutoff to remove the
region of the electron-positron phase space from which the wave
function receives most of its strength.  This requires
\begin{equation}
\Lambda^2>\alpha^2 m^2 \;.
\end{equation}

While it is not necessary, the most elegant way to proceed is to
introduce `new' variables,
\begin{equation}
\kappa_i = k_{\perp i} \;,
\end{equation}
\begin{equation}
y_i={1 \over 2}+{k_z \over 2 \sqrt{{\bf k}_\perp^2 +k_z^2+m^2}} \;.
\end{equation}
This change of variables can be `discovered' in a number of ways, but
they basically take us back to equal time coordinates, in which both
boost and rotational symmetries are kinematic after a nonrelativistic
reduction.

For cutoffs that satisfy $\alpha m^2 > \Lambda^2 > \alpha^2 m^2$, Eq.
(145) simplifies tremendously when all terms of higher order
than $\alpha^2$ are dropped.  Using the scaling behavior of the
momenta, and the fact that we will find $E^2$ is
$\order(\alpha^4)$, Eq. (145) reduces to:
\begin{equation}
-E \phi({\bf k}_1) = { {\bf k}_1^2 \over m} \phi({\bf k}_1)
-\alpha \int {d^3 k_2 \over (2 \pi)^3} {1 \over \bigl({\bf k}_1^2 -
{\bf k}_2^2\bigr)} \phi({\bf k}_2) \;.
\end{equation}
The step function cutoffs drop out to leading order, leaving us with
the familiar nonrelativistic Schr{\"o}dinger equation for positronium
in momentum space.  The solution is
\begin{equation}
\phi({\bf k}) = {{\cal N} \over \bigl({\bf k}^2+m E\bigr)^2} \;,
\end{equation}
\begin{equation}
E = {1 \over 4} \alpha^2 m \;.
\end{equation}
${\cal N}$ is a normalization constant.

This is the Bohr energy for the ground state of positronium, and it
is obvious that the entire nonrelativistic spectrum is reproduced to
leading order.

Beyond this leading order result the calculations become much more
interesting, and in any Hamiltonian formulation they rapidly become
complicated.  There is no analytic expansion of the binding energy in
powers of $\alpha$, since negative mass-squared states appear to
signal vacuum instability when $\alpha$ is negative ; but our simple
strategy for performing field theory calculations can be improved by
taking advantage of the weak coupling expansion.\site{nonpert}

We can expand the binding energy in powers of $\alpha$, and as is
well known we find that powers of ${\rm log}(\alpha)$ appear in the
expansion at $\order(\alpha^5)$.  We have taken the first step to
generate this expansion by expanding the effective Hamiltonian in the
explicit powers of $\alpha$ which appear in the renormalization group
analysis.  The next step is to take advantage of the fact that all
bound state momenta are proportional to $\alpha$, which allows us to
expand each of the operators in the Hamiltonian in powers of
momenta.  The renormalization group analysis justifies an expansion
in powers of transverse momenta, and the nonrelativistic reduction
leads to an expansion in powers of longitudinal momentum
differences. The final step is to regroup terms appearing in bound
state perturbation theory.  For example, when we compute the first
order correction in bound state perturbation theory, we find all
powers of $\alpha$, and these must be grouped order-by-order with
terms that appear at higher orders of bound state perturbation theory.

The leading correction to the binding energy is $\order(\alpha^4)$,
and producing these corrections is a much more serious test of the
renormalization procedure than the calculation shown above. To what
order in the coupling must the Hamiltonian be computed to correctly
reproduce all masses to $\order(\alpha^4)$?  The leading error can be
found in the electron mass itself.  With the Hamiltonian given above,
two-loop effects would show errors in the electron mass that are
$\order(\alpha^2 \Lambda^2)$.  This would appear to present a problem
for the calculation of the binding energy to $\order(\alpha^2)$, but
remembering that the cutoff must be lowered so that $\alpha m^2 >
\Lambda^2 > \alpha^2 m^2$, we see that the error in the electron mass
is actually of $\order(\alpha^{3+\delta} m^2)$.  This means that to
compute masses correctly to $\order(\alpha^4)$ we would have to
compute the Hamiltonian to $\order(\alpha^4)$, which requires a
fourth-order similarity calculation for QED.  Such a calculation has
not yet been completed.  However, if we compute the splitting between
bound state levels instead, errors in the electron mass cancel and we
find that the Hamiltonian computed to $\order(\alpha^2)$ is
sufficient.

In Ref. \cite{BJ97a} we have shown that the fine structure of
positronium is correctly reproduced when the first- and second-order
corrections from bound state perturbation theory are added. 
This is a formidable calculation, because the exact Coulomb bound and
scattering states appear in second-order bound state perturbation
theory\footnote{There is a trick which allows this calculation to be
performed using only first-order bound state perturbation
theory.\site{Br96b} The trick basically involves using a Melosh
rotation.}

A complete calculation of the Lamb shift in hydrogen would also
require a fourth-order similarity calculation of the Hamiltonian;
however, the dominant contribution to the Lamb shift that was first
computed by Bethe\site{bethe} can be computed using a Hamiltonian
determined to $\order({\alpha})$.\site{BJ97c}  In this calculation a
Bloch transformation was used rather than a similarity transformation
because the Bloch transformation is simpler and small energy
denominator problems can be avoided in analytic QED calculations.

The primary obstacle to using our light-front strategy for precision
QED calculations is algebraic complexity.  We have successfully used
QED as a testing ground for this strategy, but these calculations can
be done much more conveniently using other methods.  The theory for
which we believe our methods are best suited is QCD.

\section{LIGHT-FRONT QCD}

This section relies heavily on the discussion of positronium, because
we only require the QCD Hamiltonian determined to $\order(\alpha)$ to
discuss a simple confinement mechanism which appears naturally in
light-front QCD and to complete reasonable zeroth order calculations
for heavy quark bound states.  To this order the QCD Hamiltonian in
the quark-antiquark sector is almost identical to the QED Hamiltonian
in the electron-positron sector.  Of course the QCD Hamiltonian
differs significantly from the QED Hamiltonian in other sectors, and
this is essential for justifying my choice of $\H_0$ for
non-perturbative calculations.

The basic strategy for doing a sequence of (hopefully) increasingly
accurate QCD bound state calculations is almost identical to the
strategy for doing QED calculations.  I use coupling coherence to
find an expansion for $H^\Lambda$ in powers of the QCD coupling
constant to a finite order.  I then divide the Hamiltonian into a
non-perturbative part, $\h0$, and a perturbative part, $\V$.  The
division is based on the physical argument that adding a parton in an
intermediate state should require more energy than indicated by the
free Hamiltonian, and that as a result these states will `freeze out'
as the cutoff approaches $\Lambda_{QCD}$. When this happens the
evolution of the Hamiltonian as the cutoff is lowered further changes
qualitatively, and operators that were consistently canceled over an
infinite number of scales also freeze, so that their effects in the
few parton sectors can be studied directly.  A one-body operator and
a two-body operator arise in this fashion, and serve to confine both
quarks and gluons.

The simple confinement mechanism I outline is certainly not the final
story, but it may be the seed for the full confinement mechanism.  One
of the most serious problems we face when looking for non-perturbative
effects such as confinement is that the search itself depends on the
effect.  A candidate mechanism must be found and then shown to
self-consistently produce itself as the cutoff is lowered towards
$\Lambda_{QCD}$.

Once we find a candidate confinement mechanism, it is possible
to study heavy quark bound states with little modification of the QED
strategy.  Of course the results in QCD will differ from those in
QED because of the new choice of $\H_0$, and in higher orders
because of the gluon interactions.  When we move on to light quark
bound states, it becomes essential to introduce a mechanism for chiral
symmetry breaking.\site{mustaki,nonpert} I will discuss this briefly
at the end of this section.

When we compute the QCD Hamiltonian to $\order(\alpha)$, several
significant new features appear.  First are the familiar gluon
interactions.  In addition to the many gluon interactions found in the
canonical Hamiltonian, there are modifications to the instantaneous
gluon exchange interactions, just as there were modifications to the
electron-positron interaction.  For example, a Coulomb interaction
will automatically arise at short distances.  In addition the gluon
self-energy differs drastically from the photon self-energy.

The photon develops a self-energy because it mixes with
electron-positron pairs, and this self energy is $\order(\alpha
\Lambda^2/P^+)$.  When the cutoff is lowered below $4 m^2$, this mass
term vanishes because it is no longer possible to produce
electron-positron pairs.  For all cutoffs the small bare
photon self-energy is exactly canceled by mixing
with pairs below the cutoff.  I will not go through the
calculation, but because the gluon also mixes with gluon pairs in
QCD, the gluon self-energy acquires an infrared divergence, just as
the electron did in QED.  In QCD both the quark and gluon
self-energies are proportional to $\alpha \Lambda^2
\ln(1/\epsilon)/P^+$, where $\epsilon$ is the secondary cutoff on
parton longitudinal momenta introduced in the last section.  This
means that even when the primary cutoff $\Lambda^2$ is finite, the
energy of a single quark or a single gluon is infinite, because we are
supposed to let $\epsilon \rightarrow 0$.  One can easily argue that
this result is meaningless, because the relevant matrix elements of
the Hamiltonian are not even gauge invariant; however, since we must
live with a variational principle when doing Hamiltonian
calculations, this result may be useful.

In QED I argued that the bare electron self-energy was a complete lie,
because the bare electron mixes with photons carrying arbitrarily small
longitudinal momenta to cancel this bare self-energy and produce a
finite mass physical electron.  However, in QCD there is no reason to
believe that this perturbative mixing continues to arbitrarily small
cutoffs.  There are {\it no} massless gluons in the world. In this
case, the free QCD Hamiltonian is a complete lie and {\it cannot be
trusted} at low energies.

On the other hand, coupling coherence gives us no choice about the
quark and gluon self-energies as computed in perturbation theory.
These self-energies appear because of the behavior of the theory at
extremely high energies.  The question is not whether large
self-energies appear in the Hamiltonian.  The question is whether
these self-energies are canceled by mixing with low energy
multi-gluon states.  I argue that this cancellation does not occur,
and that the infrared divergent quark and gluon self-energies
{\it should be included} in $\H_0$.  The transverse scale for these energies
is the running scale $\Lambda$, and over many orders of magnitude we
should see the self-energies canceled by mixing.  However, as the
cutoff approaches $\Lambda_{QCD}$, I speculate that these
cancellations cease to occur because perturbation theory breaks
down and a mass gap between states with and without extra gluons
appears.

But if the quark and gluon self-energies diverge, and the divergences
cannot be canceled by mixing between sectors with an increasingly
large number of partons, how is it possible to obtain finite mass
hadrons?  The parton-parton interaction also diverges, and the
infrared divergence in the two-body interaction {\it exactly cancels} the
infrared divergence in the one-body operator for color singlet
states.

Of course, the cancellation of infrared divergences is not enough to
obtain confinement.  The cancellation is exact regardless of the
relative motion of the partons in a color singlet state, and
confinement requires a residual interaction.  I will show that the
$\order(\alpha)$ QCD Hamiltonian produces a logarithmic potential in
both longitudinal and transverse directions.  I will not discuss
whether a logarithmic confining potential is `correct,' but to the
best of my knowledge there is no rigorous demonstration that the
confining interaction is linear, and a logarithmic potential
may certainly be of interest phenomenologically for heavy quark bound
states.\site{quigg, quigg97} I would certainly be delighted if a
better light-front calculation produces a linear potential, but this
may not be necessary even for light hadron calculations.

The calculation of how the quark self-energy changes when a similarity
transformation lowers the cutoff on energy transfer is almost
identical to the electron self-energy calculation. Following the steps
in the section on positronium, we find the one-body operator required
by coupling coherence,
\begin{eqnarray}
\Sigma^{\Lambda}_{coh}(p)&=&
{g^2 C_F \over 8\pi^2 p^+} \Biggl\{2 y \Lambda^2 \ln\Biggl({ y^2
\Lambda^2 \over (y \Lambda^2+m^2) \epsilon }
\Biggr) -{3 \over 2} y
\Lambda^2+{1 \over 2} {y m^2 \Lambda^2 \over y \Lambda^2+m^2} \nonumber \\
&&~~~~~~~~~~~~~~~+ 3 m^2
\ln\Biggl( {m^2 \over y \Lambda^2 + m^2} \Biggr) \Biggl\} + {\cal
O}(\epsilon/y) \;,
\end{eqnarray}
where $C_F=(N^2-1)/(2N)$ for a SU(N) gauge theory.

The calculation of the quark-antiquark interaction required
by coupling coherence is also nearly identical to the QED
calculation.  Keeping only the infrared singular parts of the
interaction, as was done in QED,
\begin{eqnarray}
\tilde{V}_{coh}^{\Lambda} &=&
 - 4 g_{\Lambda}^2 C_F \sqrt{p_1^+ p_2^+ k_1^+ k_2^+}
{q_\perp^2 \over (q^+)^3}  \delta_{\sigma_1 \sigma_2}
\delta_{\lambda_1 \lambda_2} \nonumber \\
&&~~~~~~~\theta\bigl(|q^+|-\epsilon P^+\bigr)~
\theta\Biggl(\Lam-\mid p_1^-+k_1^--p_2^--k_2^- \mid
\Biggr) \nonumber \\
&&~~\Biggl[ {\theta\bigl(|p_1^- -p_2^- -q^-| -\Lambda^2 / {\cal P}^+
\bigr) \;\; \theta\bigl(|p_1^- -p_2^- -q^-|- |k_2^- -k_1^- -q^-|
\bigr) \over p_1^- -p_2^- -q^-} \nonumber \\ &&~~~~
+{\theta\bigl(|k_2^- -k_1^- -q^-| - \Lambda^2 / {\cal P}^+ \bigr)
\;\; \theta\bigl( |k_2^- -k_1^- -q^-| - |p_1^- -p_2^- -q^-| \bigr)
\over k_2^- -k_1^- -q^-} \Biggr]  \;. \nonumber\\
\end{eqnarray}

The instantaneous gluon exchange interaction is
\begin{eqnarray}
V^\Lambda_{instant} &=& 
- 4 g_{\Lambda}^2 C_F \sqrt{p_1^+ p_2^+ k_1^+ k_2^+}
\Biggl({1 \over q^+}\Biggr)^2 \delta_{\sigma_1 \sigma_2}
\delta_{\lambda_1 \lambda_2} \nonumber \\
&&~~\times \theta\bigl(|q^+|-\epsilon P^+\bigr)~
\theta\Biggl(\Lam-\mid p_1^-+k_1^--p_2^--k_2^- \mid
\Biggr)
\;.
\end{eqnarray}
Just as in QED the coupling coherent interaction induced by
gluon exchange above the cutoff partially cancels instantaneous gluon
exchange.  For the discussion of confinement the part of $V_{coh}$
that remains is not important, because it produces the short
range part of the Coulomb interaction.  However, the part of the
instantaneous interaction that is not canceled is
\begin{eqnarray}
\tilde{V}^\Lambda_{instant} &=& - 4 g_{\Lambda}^2 C_F
\sqrt{p_1^+ p_2^+ k_1^+ k_2^+}
\Biggl({1 \over q^+}\Biggr)^2 \delta_{\sigma_1 \sigma_2}
\delta_{\lambda_1 \lambda_2} \nonumber \\
&&\times
\theta\Biggl(\Lam-\mid p_1^-+k_1^--p_2^--k_2^- \mid \Biggr)
\theta\bigl(|p_1^+-p_2^+|-\epsilon P^+\bigr) \nonumber \\
&&\times \Biggl[ \theta\bigl(\Lam -|p_1^- -p_2^- -q^-|  \bigr)
\;\; \theta\bigl(|p_1^- -p_2^- -q^-|- |k_2^- -k_1^- -q^-| \bigr) +
\nonumber \\
&&\theta\bigl(\Lam-|k_2^- -k_1^- -q^-| \bigr)
\;\; \theta\bigl( |k_2^- -k_1^- -q^-| - |p_1^- -p_2^- -q^-| \bigr)
\Biggr] \;.
\end{eqnarray}

Note that this interaction contains a cutoff that projects onto
exchange energies below the cutoff, because the interaction has been
screened by gluon exchange above the cutoffs. This interaction can
become important at long distances, if parton exchange below the
cutoff is dynamically suppressed.  In QED I argued that this singular
long range interaction is exactly canceled by photon exchange below
the cutoff, because such exchange is not suppressed no matter how low
the cutoff becomes.  Photons are massless and experience no
significant interactions, so they are exchanged to arbitrarily low
energies as effectively free photons.  This cannot be the case for
gluons.

For the discussion of confinement, I will place only the most singular
parts of the quark self-energy and the quark-antiquark interaction in
$\h0$.  To see that all infrared divergences cancel and that the
residual long range interaction is logarithmic, we can study the
matrix element of these operators for a quark-antiquark state,
\begin{eqnarray}
|\Psi(P)\rangle &=& \sum_{\sigma \lambda} \sum_{rs}
\int {dp^+ d^2p_\perp \over
16\pi^3 p^+} {dk^+ d^2k_\perp \over 16\pi^3 k^+} \sqrt{p^+ k^+}
16\pi^3 \delta^3(P-p-k) \nonumber \\
&&~~~~~~~~~~~~~~\phi(p,\sigma,r;k,\lambda,s) b^{r\dagger}(p,\sigma)
d^{s\dagger}(k,\lambda) |0\rangle \;,
\end{eqnarray}
where $r$ and $s$ are color indices and I will choose
$\phi$ to be a color singlet and drop color indices.  The
cancellations we find do not occur for the color octet
configuration.  The matrix element is,
\begin{eqnarray}
\langle \Psi(P)|\H_0|\Psi(P')\rangle &=& 16\pi^3 P^+ \delta^3(P-P')
\times \nonumber \\
&\Biggl\{& \int {dy d^2\kappa \over 16\pi^3}
\Biggl[{g_\Lambda^2 C_F \Lambda^2 \over 2\pi^2 P^+} \; \ln\bigl({1
\over \epsilon}\bigr) \Biggr] |\phi(\kappa,y)|^2 \nonumber \\
&-&{4 g_\Lambda^2 C_F \over P^+} \int {dy_1 d^2\kappa_1 \over 16\pi^3}
{dy_2 d^2\kappa_2 \over 16\pi^3} \theta\Biggl(\Lambda^2-\mid
{\kappa_1^2+m^2 \over y_1(1-y_1)}-{\kappa_2^2+m^2 \over y_2(1-y_2)}
\mid\Biggr) \nonumber \\
&&~~\times \Biggl[
\theta\Biggl(\Lambda^2- \mid {\kappa_1^2+m^2 \over y_1} - {\kappa_2^2
+m^2 \over y_2}-{(\kappa_1-\kappa_2)^2 \over |y_1-y_2|}\mid \Biggr)
\nonumber \\ &&~~~~~~
\times \theta\Biggl(\mid {\kappa_1^2+m^2 \over y_1} - {\kappa_2^2
+m^2 \over y_2}-{(\kappa_1-\kappa_2)^2 \over |y_1-y_2|}\mid -
\nonumber \\ &&~~~~~~~~~~~~~~~~~~~~~~~~~~
\mid {\kappa_2^2+m^2 \over 1-y_2} - {\kappa_1^2 +m^2
\over 1-y_1}-{(\kappa_1-\kappa_2)^2 \over |y_1-y_2|}
\mid\Biggr)  \nonumber \\ && ~~~~~~ +
\theta\Biggl(\Lambda^2-\mid {\kappa_2^2+m^2 \over 1-y_2} -
{\kappa_1^2 +m^2 \over 1-y_1}-{(\kappa_1-\kappa_2)^2 \over |y_1-y_2|}
\mid\Biggr)  \nonumber \\ && ~~~~~~
\times \theta\Biggl(\mid {\kappa_2^2+m^2 \over 1-y_2} - {\kappa_1^2
+m^2 \over 1-y_1}-{(\kappa_1-\kappa_2)^2 \over |y_1-y_2|}\mid -
\nonumber \\ &&~~~~~~~~~~~~~~~~~~~~~~~~~~
\mid {\kappa_1^2+m^2 \over y_1} - {\kappa_2^2
+m^2 \over y_2}-{(\kappa_1-\kappa_2)^2 \over |y_1-y_2|}\mid \Biggr)
\Biggr]
\nonumber \\ &&~~
\times \theta\Bigl(\mid y_1-y_2 \mid -\epsilon\Bigr)
\Biggl({1 \over y_1-y_2 }\Biggr)^2 \phi^*(\kappa_2,y_2)
\phi(\kappa_1,y_1)
\Biggr\}.
\end{eqnarray}
Here I have chosen a frame in which the center-of-mass
transverse momentum is zero, assumed that the longitudinal momentum
scale introduced by the cutoffs is that of the bound state, and used
Jacobi coordinates,
\begin{equation}
p_i^+=y_i P^+\;,\;\;p_{i\perp}=\kappa_i\;\;;\;\;\; k_i^+=(1-y_i)P^+\;,
\;\; k_{i\perp}=-\kappa_i \;.
\end{equation}

The first thing I want to do is show that the last term is divergent
and the divergence exactly cancels the first term.  My demonstration
is not elegant, but it is straightforward.  The divergence results
from the region $y_1 \sim y_2$.  In this region the second and third
cutoffs restrict $(\kappa_1-\kappa_2)^2$ to be small compared to
$\Lambda^2$, so we should change variables,
\begin{equation}
Q={\kappa_1+\kappa_2 \over 2} \;,\;\;Y={y_1+y_2 \over 2} \;\;;\;\;\;
q=\kappa_1-\kappa_2 \;,\;\;y=y_1-y_2 \;.
\end{equation}
Using these variables we can approximate the above
interaction near $q=0$ and $y=0$.  The double integral becomes
\begin{eqnarray}
{-4 g_\Lambda^2 C_F \over P^+} \int {dY d^2Q \over 16\pi^3} {dy d^2q
\over 16\pi^3} \theta(1-Y) \theta(Y) |\phi(Q,Y)|^2 \nonumber \\
~~~~~\theta\Biggl(\Lambda^2-{q^2 \over |y|}\Biggr) \theta\bigl(
|y|-\epsilon\bigr) \theta\bigl(\eta-|y|\bigr)
~ \Biggl({1 \over y}\Biggr)^2 \;,
\end{eqnarray}
where $\eta$ is an arbitrary constant that restricts $|y|$
from becoming large. Completing the $q$ and $y$ integration we get
\begin{equation}
-{g_\Lambda^2 C_F \Lambda^2 \over 2\pi^2 P^+} \ln\Biggl({1 \over
\epsilon}\Biggr) \int {dY d^2Q \over 16\pi^3}
\theta(1-Y) \theta(Y) |\phi(Q,Y)|^2 \;.
\end{equation}

The divergent part of this exactly cancels the first term on the
right-hand side of Eq. (122).  This cancellation occurs for any state,
and this cancellation is unusual because it is between the expectation
value of a one-body operator and the expectation value of a two-body
operator.  The cancellation resembles what happens in the Schwinger
model and is easily understood.  It results from the fact that a color
singlet has no color monopole moment.  If the state is a color octet
the divergences are both positive and cannot cancel.  Since the
cancellation occurs in the matrix element, {we can let $\epsilon
\rightarrow 0$ before diagonalizing} $\H_0$.

The fact that the divergences cancel exactly does not indicate that
confinement occurs.  This requires the residual interactions to
diverge at large distances, which means small momentum transfer.
Equivalently, we need the color dipole self-energy to diverge if the
color dipole moment diverges because the partons separate to large
distance.

My analysis of the residual interaction is neither elegant nor
complete.  For a more complete analysis see Ref.\cite{Br96a}. I 
show that the interaction is logarithmic in the longitudinal
direction at zero transverse separation and logarithmic in the
transverse direction at zero longitudinal separation, and I
present the full angle-averaged interaction without derivation.

In order to avoid the infrared divergence, which is canceled, I
compute spatial derivatives of the potential.
First consider the potential in the longitudinal direction.
Given a momentum space expression, set $x_\perp=0$ and the
Fourier transform of the longitudinal interaction requires the
transverse momentum integral of the potential,
\begin{equation}
{\partial \over \partial z} V(z) = \int {d^3 q \over (2\pi)^3} i q_z
V(q_\perp,q_z) e^{i q_z z} \;.
\end{equation}
We are interested only in the long range potential, so we
can assume that $q_z$ is arbitrarily small during the analysis and
approximate the step functions accordingly.

For our interaction this leads to
\begin{eqnarray}
{\partial \over \partial x^-} V(x^-) &=& -4 g_\Lambda^2 C_F P^+ \int
{dq^+ d^2q_\perp \over 16\pi^3} \theta\bigl(P^+-|q^+|\bigr) \nonumber
\\ &&~~~~~\theta\bigl(\Lam-{q_\perp^2 \over |q^+|}\bigr) \Bigl( {1
\over q^+} \Bigr)^2 (i q^+) e^{i q^+ x^-} \;.
\end{eqnarray}
Completing the $q_\perp$ integration we have
\begin{eqnarray}
{\partial \over \partial x^-} V(x^-) &=&  - {i g_\Lambda^2 C_F
\Lambda^2 \over 4\pi^2} \int dq^+ \theta\bigl(P^+-|q^+|\bigr) {q^+
\over |q^+|} e^{i q^+ x^-} \nonumber \\
&=& {g_\Lambda^2 C_F \Lambda^2 \over 2\pi^2} \int_0^{P^+} dq^+
\sin\bigl({q^+ x^-}\bigr) \nonumber \\
&=& {g_\Lambda^2 C_F \Lambda^2 \over 2\pi^2} \Biggl( {1 \over x^-} -
{\cos\bigl( P^+ x^- \bigr) \over x^-}\Biggr) \nonumber \\
&=& {g_\Lambda^2 C_F \Lambda^2 \over 2\pi^2}  {\partial \over
\partial x^-} \;
\ln\bigl(|x^-|) + {\rm short~range}\;.
\end{eqnarray}
To see that the term involving a cosine in the next-to-last
line produces a short range potential, simply integrate it.  At large
$|x^-|$, which is the only place we can trust our approximations for
the original integrand, this yields a logarithmic potential, as
promised.

Next consider the potential in the transverse direction.  Here we can
set $x^-=0$ and get
\begin{eqnarray}
{\partial \over \partial x_\perp^i} V(x_\perp) &=& -4 g_\Lambda^2 C_F
P^+ \int {dq^+ d^2 q_\perp \over 16\pi^3} \theta\bigl(P^+-|q^+|\bigr)
\nonumber \\ &&~~~~~\theta\Biggl(\Lam-{q_\perp^2 \over |q^+|}\Biggr)
\Biggl( {1 \over q^+}\Biggr)^2 (i q_\perp^i) e^{i {\bf q}_\perp \cdot
{\bf x}_\perp} \;.
\end{eqnarray}
Here I have used the fact that the integration is dominated
by small $q^+$ to simplify the integrand again.  Completing the $q^+$
integration this becomes
\begin{eqnarray}
{\partial \over \partial x_\perp^i} V(x_\perp) &=& -{i g_\Lambda^2 C_F
\Lambda^2 \over 2\pi^3} \int d^2q_\perp {q_\perp^i \over q_\perp^2}
e^{ i {\bf q}_\perp \cdot {\bf x}_\perp}~+~{\rm short~range}\nonumber \\
&=& {g_\Lambda^2 C_F \Lambda^2 \over \pi^2} {x_\perp^i \over x_\perp^2}
\nonumber \\
&=& {g_\Lambda^2 C_F \Lambda^2 \over \pi^2} {\partial \over \partial
x_\perp^i} \;\ln(|{\bf x_\perp}|) \;.
\end{eqnarray}
Once again, this is the derivative of a logarithmic
potential, as promised.  The strength of the long-range logarithmic
potential is not spherically symmetrical in these coordinates, with
the potential being larger in the transverse than in the longitudinal
direction.  Of course, there is no reason to demand that the
potential is rotationally symmetric in these coordinates.

The angle-averaged potential for two heavy quarks with
mass $M$ is\site{Br96a}
\begin{equation}
{g_\Lambda^2 C_F \Lambda^2 \over \pi^2} \Bigl(\ln({\cal R}) -
{\rm Ci}({\cal R}) + 2 {{\rm Si}({\cal R}) \over {\cal R}} - {(1- {\rm
cos}({\cal R})) \over {\cal R}^2} + {{\rm sin}({\cal R}) \over {\cal
R}} - {5
\over 2}+\gamma_E\Bigr) \;,
\end{equation}
where,
\begin{equation}
{\cal R} = {\Lambda^2 \over 2M} r \;.
\end{equation}
I have assumed the longitudinal momentum scale in the cutoff equals
the longitudinal momentum of the state.

The full potential is not naively rotationally invariant.  Of course,
the two body interaction in QED is also not rotationally invariant. 
The leading term in an expansion in powers of momenta yields the
rotationally invariant Coulomb interaction, but higher order terms
are not rotationally invariant.  These higher order terms do not
spoil rotational invariance of physical results because they must be
combined with higher order interactions in the effective Hamiltonian,
and with corrections involving photons at higher orders in bound
state perturbation theory.  There is no reason to expect, and no real
need for a potential that is rotationally invariant; however, to
proceed we need to decide which part of the potential must be treated
non-perturbatively.  The simplest reasonable choice is the
angle-averaged potential, which is what we have used in heavy quark
calculations.\site{Br96a,Br97a}

Had we computed the quark-gluon or gluon-gluon interaction, we would
find essentially the same residual long range two-body interaction in
every Fock space sector, although the strengths would differ because
different color operators appear.  In QCD gluons have a
divergent self-energy and experience divergent long range
interactions with other partons if we use coupling coherence.  In
this sense, the assumption that gluon exchange below some cutoff is
suppressed is consistent with the Hamiltonian that results from this
assumption.  To show that gluon exchange is suppressed when $\Lambda
\rightarrow \Lambda_{QCD}$, rather than some other scale ({\it i.e.},
zero as in QED), a non-perturbative calculation of gluon exchange is
required.  This is exactly the calculation bound state perturbation
theory produces, and bound state perturbation theory suggests how the
perturbative renormalization group calculation might be modified to
generate these confining interactions self-consistently.

If perturbation theory, which produced this potential, continues to be
reasonable, the long range potential will be exactly canceled in QCD
just as it is in QED. We need this exact cancellation of new forces to
occur at short distances and turn off at long distances, if we want
asymptotic freedom to give way to a simple constituent confinement
mechanism. At short distances the divergent self-energies and two-body
interactions cannot be ignored, but they should exactly cancel
pairwise if these divergences appear only when gluons are emitted and
absorbed in immediately successive vertices, as I have speculated.
The residual interaction must be analyzed more carefully at short
distances, but in any case a logarithmic potential is less singular
than a Coulomb interaction, which does not disturb asymptotic
freedom.  This is the easy part of the problem.  The hard part is
discovering how the potential can survive at any scale.

A perturbative renormalization group will not solve this problem.  The
key I have suggested is that interactions between quarks and gluons in
the intermediate states required for cancellation of the potential
will eventually produce a non-negligible energy gap.  I am unable to
detail this mechanism without an explicit calculation, but let me
sketch a naive picture of how this might happen.

Focus on the quark-antiquark-gluon intermediate state, which mixes
with the quark-antiquark state to screen the long range potential.
The free energy of this intermediate state is always higher than that
of the quark-antiquark free energy, as is shown using a simple
kinematic argument.\site{nonpert} However, if the gluon is massless,
the energy gap between such states can be made arbitrarily small.  As
we use a similarity transformation to run a vertex cutoff on energy
transfer, mixing persists to arbitrarily small cutoffs since the gap
can be made arbitrarily small.  Wilson has suggested using a gluon
mass to produce a gap that will prevent this mixing from persisting as
the cutoff approaches\site{nonpert} $\Lambda_{QCD}$, and I am
suggesting a slightly different mechanism.

If we allow two-body interactions to act in both sectors to all
orders, even the Coulomb interaction can produce quark-antiquark and
quark-antiquark-gluon bound states.  In this respect QCD again differs
qualitatively from QED because the photon cannot be bound and the
energy gap is always arbitrarily small even when the electron-positron
interaction produces bound states.  If we assume that a fixed energy
gap arises between the quark-antiquark bound states and the
quark-antiquark-gluon bound states, and that this gap establishes the
important scale for non-perturbative QCD, these states must cease to
mix as the cutoff goes below the gap.

An important ingredient for any calculation that tries to establish
confinement self-consistently is a {\it seed mechanism}, because it is
possible that it is the confining interaction itself which alters the
evolution of the Hamiltonian so that confinement can arise.  I have
proposed a simple seed mechanism whose perturbative origin is
appealing because this allows the non-perturbative evolution induced
by confinement to be matched on to the perturbative evolution required
by asymptotic freedom.

I provide only a brief summary of our heavy quark bound
state calculations,\site{Br96a,Br97a} and refer the reader to the
original articles for details.  We follow the strategy that has been
successfully applied to QED, with modifications suggested by the fact
that gluons experience a confining interaction.  We keep only the
angle-averaged two-body interaction in $\H_0$, so the zeroth order
calculation only requires the Hamiltonian matrix elements in the
quark-antiquark sector to $\order(\alpha)$.  All of the matrix
elements are given above.

For heavy quark bound states\site{Br97a} we can simplify the
Hamiltonian by making a nonrelativistic reduction and solving a
Schr{\"o}dinger equation, using the potential in Eq. (168).  We must
then choose values for $\Lambda$, $\alpha$, and $M$.  These should be
chosen differently for bottomonium and charmonium. The cutoff for
which the constituent approximation works well depends on the
constituent mass, as in QED where it is obviously different for
positronium and muonium.  In order to fit the ground state and first
two excited states of charmonium, we use $\Lambda=2.5 GeV$,
$\alpha=0.53$, $M_c=1.6 GeV$.  In order to fit these states in
bottomonium we use $\Lambda=4.9 GeV$, $\alpha=0.4$, and $M_b=4.8
GeV$.\footnote{There are several minor errors in Ref. \cite{Br97a},
which are discussed in Ref.\cite{Sz97a}.  I have also chosen ${\cal
P}^+=P^+$, and in those papers ${\cal P}^+=P^+/2$; and I have absorbed
this change in a redefinition of $\Lambda$.}  Violations of
rotational invariance from the remaining parts of the potential are
only about 10\%, and we expect corrections from higher Fock state
components to be at least of this magnitude for the couplings we use.

These calculations show that the approach is reasonable, but they are
not yet very convincing.  There are a host of additional calculations
that must be done before the success of this approach can be judged,
and most of them await the solution of several technical problems. 

In order to test the constituent approximation and see that full
rotational invariance is emerging as higher Fock states enter the
calculation, we must be able to include gluons.  If gluons are
massless (which need not be true since the gluon mass is a relevant
operator that must in principle be tuned), we cannot continue to
employ a nonrelativistic reduction.  In any case, we are primarily
interested in light hadrons and must learn how to perform
relativistic calculations to study them.  The primary difficulty is
that evaluation of matrix elements of the interactions given above
involves high dimensional integrals which display little symmetry in
the presence of cutoffs.  Worse than this is the fact that the
confinement mechanism requires cancellation of infrared divergences
which have prevented us from using Monte Carlo methods to
date.\site{BA97a} These difficulties are avoided when a
nonrelativistic reduction is made, but there is little more that we
can do for which such a reduction is valid.

I conclude this section with a few comments on chiral symmetry
breaking.  While quark-quark, quark-gluon, and gluon-gluon confining
interactions  appear in the hamiltonian at {\cal O}($\alpha$), 
chiral symmetry is not broken at any finite order in the 
perturbative expansion; so symmetry-breaking operators must be
allowed  to appear {\it ab initio}. Light-front
chiral symmetry differs from  equal-time chiral symmetry in several
interesting and important  aspects.\site{nonpert,mustaki,Br94b} For
example, quark masses in the kinetic energy operator do not violate
light-front chiral  symmetry; and the only operator in the canonical
hamiltonian that  violates this symmetry is a quark-gluon coupling
linear in the quark  mass.  Again, the primary technical difficulty
is the need for relativistic bound state calculations, and real
progress on chiral symmetry cannot be made before we are able to
perform relativistic calculations with constituent gluons.  If
transverse locality is maintained, simple renormalization group
arguments indicate that chiral symmetry should be broken by a
relevant operator in order to preserve well-established perturbative
results at high energy.  The only relevant operator that breaks
chiral symmetry involves gluon emission and absorption, leading to
the conclusion that the pion contains significant constituent glue. 
This situation is much simpler than what we originally
envisioned,\site{nonpert} because it does not require the addition of
any operators that cannot be found in the canonical QCD hamiltonian;
and we have long known that relevant operators must be tuned to
restore symmetries.
\vfill
\subsection{Acknowledgments}
I would like to thank the staff of the Newton Institute for their
wonderful hospitality, and the Institute itself for the support that
made this exceptional school possible.  I benefitted from many
discussions at the school, but I would like to single out Pierre van
Baal both for his patient assistance with all problems and for the
many insights into QCD he has given me.  I would also like to thank
the many theorists who have helped me understand light-front QCD,
including Edsel Ammons, Matthias Burkardt, Stan G{\l}azek, Avaroth
Harindranath, Tim Walhout, Wei-Min Zhang, and Ken Wilson.  I
especially want to thank Brent Allen, Martina Brisudov{\'a}, Billy
Jones, and Sergio Szpigel; who have been responsible for most of the
recent progress in this program.  This work has been partially
supported by National Science Foundation grants PHY-9409042 and
PHY-9511923.

{\referencestyle

}
\end{document}